\documentstyle[12pt,psfig]{article}

\input epsf
\begin{document}
\title{\bf Glassy effects in the swelling/collapse dynamics
of homogeneous polymers}
\renewcommand{\thefootnote}{\fnsymbol{footnote}}
\author{Estelle Pitard
\footnote{estelle@lpm.univ-montp2.fr}
(Harvard University, USA,\\and Laboratoire
de Physique Math\'ematique et Th\'eorique,\\ Universit\'e Montpellier II, UMR 
5825, France)
\\and Jean-Philippe Bouchaud
\footnote{bouchau@spec.saclay.cea.fr} (SPEC-CEA-Saclay, France)\\}

\renewcommand{\theequation}{\arabic{equation}}
\def\be{\begin{equation}}
\def\bea{\begin{eqnarray}}
\def\ee{\end{equation}} 
\def\eea{\end{eqnarray}} 

\maketitle
\vskip 2cm
\begin{abstract}
We investigate, using numerical simulations and analytical arguments, a simple 
one dimensional model for the swelling or the collapse of a closed polymer 
chain
of size $N$, representing the dynamical evolution of a polymer in a 
$\Theta$-solvent
that is rapidly changed into a good solvent (swelling) or a bad solvent 
(collapse).
In the case of swelling, the density profile for intermediate times is 
parabolic and
expands in space as $t^{1/3}$, as predicted by a Flory-like continuum theory. 
The dynamics slows down after a time $\propto N^2$ when the chain becomes 
stretched,
and the polymer gets stuck in metastable `zig-zag' configurations, from which 
it escapes
through thermal activation. The size of the polymer in the final stages is 
found to grow
as $\sqrt{\ln t}$. In the case of collapse, the chain very quickly (after a 
time of order
unity) breaks up into clusters of monomers (`pearls'). The evolution of the 
chain then
proceeds
through a slow growth of the size of these metastable clusters, again evolving 
as the
logarithm of time. We enumerate the total number of metastable states as 
a function
of the extension of the chain, and deduce from this computation that the 
radius of the
chain should decrease as $1/\ln(\ln t)$. We 
compute the total number of metastable states with a given value of the 
energy, and find that the complexity is non zero for arbitrary low energies.
We also obtain the distribution of cluster sizes, that we compare to simple 
`cut-in-two' coalescence models. Finally, we determine the aging properties of 
the
dynamical structure. The subaging behaviour that we find is attributed to the 
tail
of the distribution at small cluster sizes, corresponding to anomalously 
`fast'
clusters (as compared to the average). We argue that this mechanism for 
subaging might
hold in other slowly coarsening systems.
\end{abstract}

\section{Introduction}

The problem of the dynamics of a single polymer chain has been addressed in 
many
ways over the past, the Rouse model being the simplest theory \cite{deGennes}.
In the case where the polymer is not a Gaussian chain, but evolves 
either towards a swollen or collapsed state, the dynamics can also be
studied using more or less elaborate scenarios, either in a phenomenological
way (like the ``sausage'' or ``necklace''
pictures in the case of the collapse to a compact state
\cite{PGG},\cite{Goldbart}), via numerics \cite{numerics} or in an analytical
way \cite{Orland}, \cite{Dawson}. Although a very common experimental
realization of such a dynamical process is  the folding or unfolding of
proteins as they reach (respectively) their native or  denatured state, it has
been so far very difficult to test the various theoretical speculations
\cite{Eaton}.

The dynamics of polymer melts, on the other hand, is much easier to study 
experimentally \cite{Melts} and can be compared to some computer simulations
\cite{Baschnagel}. This case exhibits many interesting glassy features, but 
looks quite
different from what happens to a single polymer chain. However,
one can learn a lot from these studies, since for a long enough polymer chain 
that
collapses onto itself into a compact conformation, the local environment of a 
monomer
inside a globule is very similar to the one in a melt.
It has recently been speculated that a single homopolymeric \cite{Nikolai}
or heteropolymeric chain \cite{Shakh} does exhibit some glassy dynamics
during the collapse or folding, including aging, at least in an intermediate 
regime of time before complete folding.

While it is hard to find a satisfactory picture of what happens
microscopically as a real polymer chain collapses in three dimensions 
(either because of the length of computational time for the numerics, or 
because of the
complexity of the analytical approaches involved), one can gain some intuition 
by looking
at the simpler (but less physical) case of a one-dimensional polymer chain.

In this paper, we investigate theoretically and via Monte-Carlo simulations
the dynamical behaviour of a single one-dimensional polymer, composed of 
identical monomers
that interact via {\it local} two-body interactions. Our aim is to mimic what 
happens to
a polymeric chain in a good or a bad solvent when it swells towards an 
expanded
coil, or collapses onto a compact globule. We will present some numerical 
results for both
cases, together with some approximate analytical calculations which 
rationalize
these numerical findings. For example, we show that in several situations, 
the chain constraint (which is difficult to take into account in a rigorous 
way)
can be neglected, and the polymeric chain behaves (at least in some 
intermediate
time region) as an assembly of independent particles ruled by the same type of 
dynamics.

Finally we will discuss how our results could be compared to real systems 
easier to
observe,
like polymer melts  \cite{Baschnagel}. The model we present here
is in some way a toy model for non-disordered systems where
the aggregation interactions are most important, with or without elastic
chain constraint (a realization of that being colloidal gels where glassy 
cluster patterns
appear at high concentrations \cite{Colloids}).

\section{Model and simulations}

We consider a periodic string of $N$ particles (monomers), each of which 
constrained to live on the sites of a one-dimensional lattice. The 
elasticity of the chain is enforced by the fact that successive monomers 
along the chain cannot be at a distance larger than $m$ lattice 
spacings in space. In our numerical simulations, we have chosen $m=2$.
If one allows $m=\infty$ then the system is no more an elastic chain
but an assembly of $N$ independent particles. The polymer is constrained on a 
segment
of $L$ sites, defining a  density of monomers  $\rho=N/L$.
This length $L$ is actually the measure of the extension of the chain; as we
study the dynamics of the chain, this length will evolve with
time,
and will be referred to as $L(t)$.
 We assume that the interaction is short
range, i.e., two monomers can only interact when they stand on the same site.
(The influence of  longer ranged interactions on the physical picture obtained
below will be discussed in the conclusion). We will call $M$  the number of
sites occupied by more  than one monomer, and $n_i$ is the number of monomers
sitting on site $i$. The total energy $E$ of  the chain is therefore:
\be
E= \frac{v}{2} \sum_{i=1}^M n_i (n_i -1).
\ee
The interaction parameter $v$ is positive for an excluded volume interaction,
which leads to swelling. The ground state and stationary solution of the 
dynamics is then the fully
extended chain, with an end-to-end distance $R\sim N$. For $v<0$, the 
interaction is attractive, and we study the collapse
of the chain. In this case, the corresponding lowest-lying energy state is a 
compact ``cluster''
of $N$ particles sitting on a unique site. As will be seen in the following,
this final state is actually never reached in observable times,
as the dynamics dramatically slows down with time. 

We shall choose as our initial condition a simple random walk for the chain, 
corresponding
to an equilibrium configuration in the non-interacting case $v=0$.
The typical initial size of the chain is therefore $\sim \sqrt{N}$.  

During the Monte-Carlo simulation,  we do not allow collective motion
of monomers but only individual moves.
At each step, a monomer is allowed to move to one of its
neighbouring sites as long as the distance between its new position
and the neighbouring monomers 
along the chain is $\leq m$. In addition a monomer move is accepted 
according to
a Metropolis criterion: If $\delta E$  is the energy difference between the 
new
configuration and the initial configuration, 
then this move is accepted with probability $p=\min[1,\exp(-\frac{\delta 
E}{T})]$
where $T$ is the temperature. If one monomer on site $i$ is moved to site $j$,
the energy change $\delta E$ is given by:
\be
\delta E=\frac{v}{2}\left[(n_j +1) n_j + (n_i-1)(n_i-2) -n_j (n_j-1)-n_i 
(n_i-1)\right]=
v(n_j-n_i+1).
\ee
In the following, we will often choose $v=\pm 1$, corresponding 
respectively
to swelling or to collapse. The quantities of interest are primarily based on 
the density
profile $n(x,t)=n_i(t)$, where $x=ia$ ($a$
is the lattice spacing). For example, one can study the {\it participation 
ratios}
$Y_q(t)$, defined as:
$$
Y_q(t) = \sum_{x} n(x,t)^q.
$$
Obviously, $Y_1=N$ and $E=v(Y_2-Y_1)/2$. $Y_0$ can be taken as a measure
of the size of the chain. Alternatively, we will consider the gyration radius 
$R(t)$ defined as: 
$$
R^2(t)=\frac{1}{N} \sum_{x} \ x^2 n(x,t) - \left[\frac{1}{N} \sum_{x} x \ 
n(x,t)
\right]^2.
$$
Note that for a fully stretched chain, which is the ground state of the chain, 
the maximum value of $R$ is, for $m=2$, $N/\sqrt{12} \sim 0.28 N$.

\section{Repulsive case: swelling of the chain}

\subsection{Numerical results}

We have investigated the behaviour of the gyration radius $R(t)$ for different
chain sizes and different temperatures $T$, including $T=0$. Interestingly, 
the
different results can be rescaled on top of each other by plotting $r=R/N$ as 
a
function of $\tau=t/N^2$: see Fig. 1. There is a well defined regime of time 
scales 
where one finds $r \propto \tau^{1/z}$, with $z \simeq 3$, which holds when 
$r \ll 1$. When $r$ becomes 
of order 1, the chain becomes substantially stretched, and the dynamics stops 
(at zero temperature) or becomes activated (for non zero temperatures). The 
scaling
$r \propto \tau^{1/3}$ does not hold at very short times either, but sets in
after a time $t_{c} \sim \sqrt{N}$, as can be seen 
from Fig. 2 where we have plotted $[R(t)-R(t=0)]/\sqrt{N}$ as a function of 
$(t/\sqrt{N})$
for short times and different values of $N$. The short time behaviour is linear
in time. This initial time regime is found to be independent of the chain
constraint, up to a rescaling of time. This is expected from the simple 
theories developed below. The chain constraint only starts to play a role when
$r \sim 1$, i.e. for $t \sim N^2$.

\begin{figure}[htbp]
\begin{center}
\centerline{\psfig{figure=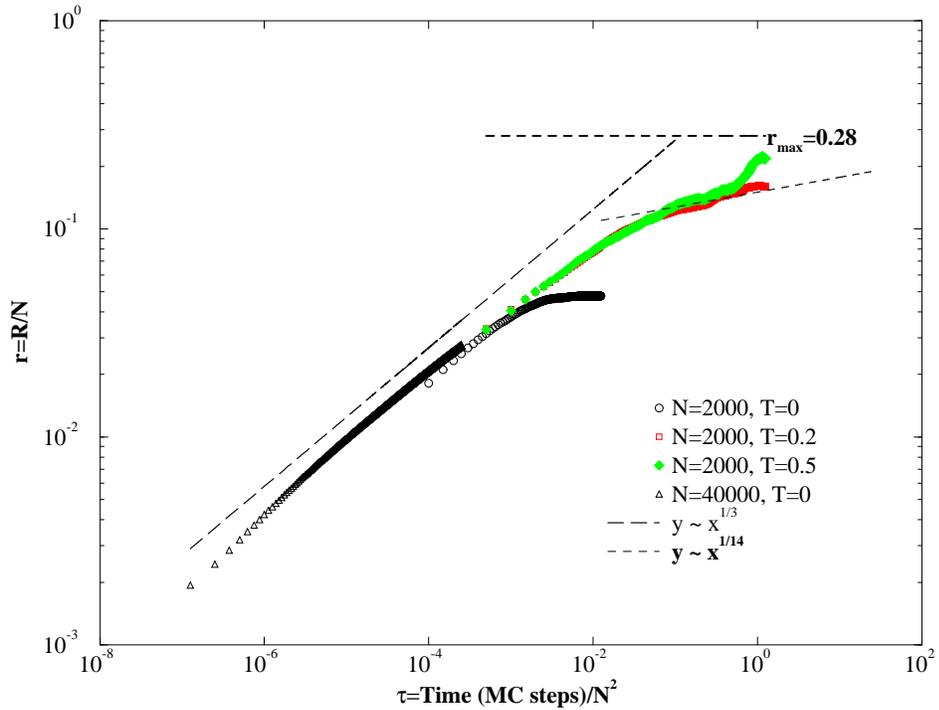,width=8cm,angle=270}}
\caption{In this graph we plot $r=R/N$ versus $\tau=t/N^2$ for various chain 
lengths and various temperatures. These are all single runs (no
averaging over initial conditions). One can see that there is a well
characterized $\tau^{1/3}$ regime for intermediate times. At zero 
temperature, the chain gets stuck in a metastable state and $r$ saturates 
to a value much smaller than its maximum value $r_{\max}=0.28$ (shown as a
dashed line). At non zero temperature, the long time regime is logarithmic in
time. We have also shown, for comparison, the $\tau^{1/14}$ behaviour
predicted by the dynamical effective length method discussed below in section
$3.2.2$.} \end{center} \end{figure}

\begin{figure}[htbp]
\begin{center}
\centerline{\psfig{figure=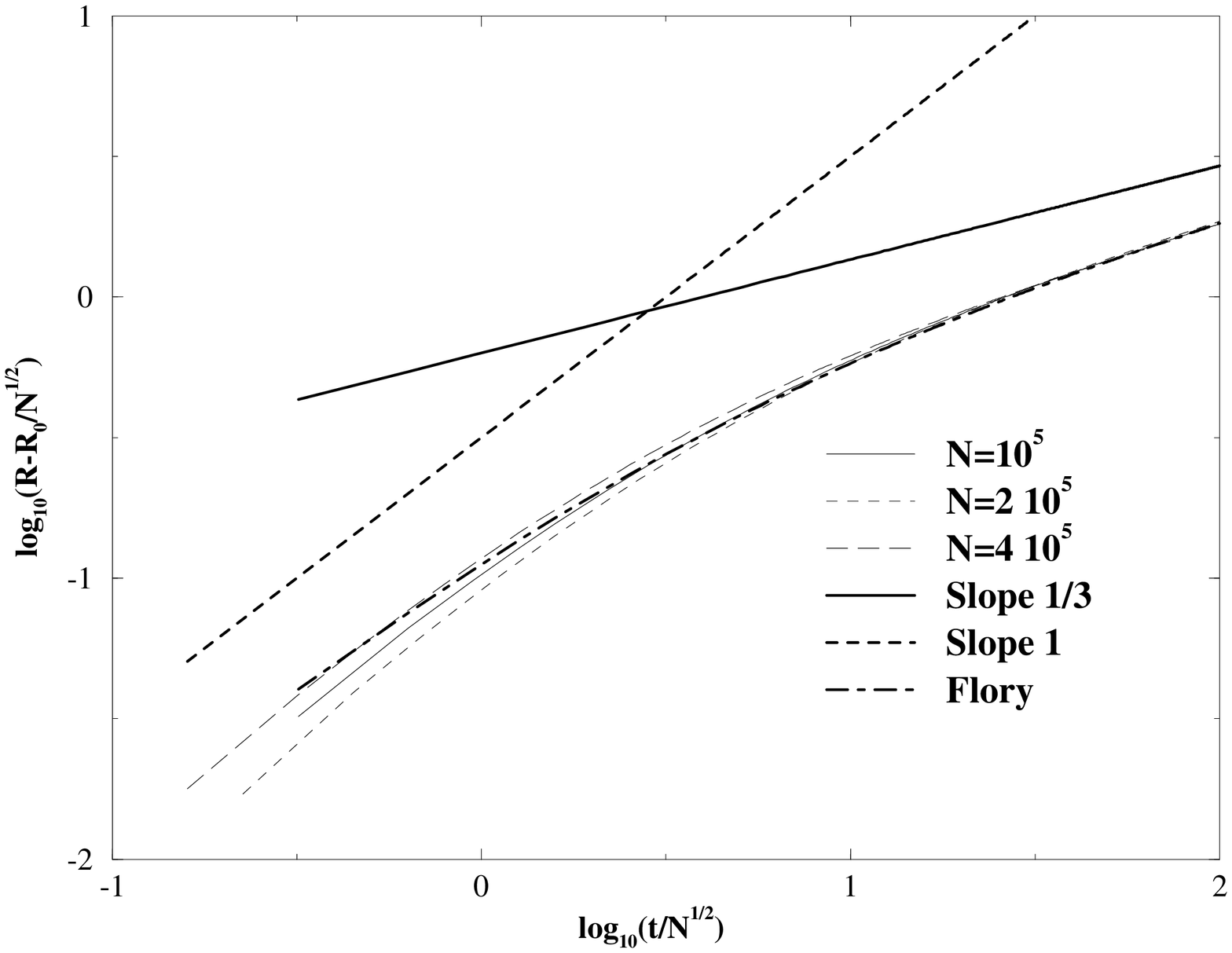,width=8cm,angle=270}}
\caption{Initial time regime, where $(R-R_0)/\sqrt{N}$ is plotted as a function
of $t/\sqrt{N}$ for different values of $N$. One sees a first regime linear in 
$t$ crossing over for $t \sim \sqrt{N}$ to the $t^{1/3}$ regime. The dot-dashed
line shows a fit to the simple Flory theory described below, Eq. 
(\protect\ref{Floryshort}).
Note that the
curves are single runs, which explains the differences for short times.}
\end{center}
\end{figure}

The dynamics of the whole density profile $n(x,t)$ from the initial 
configuration is also
interesting: after a short transient time $t_c$ ($< 100$ MC steps for $N=10 
000$) the
profile $n(x,t)$ smoothes to a parabolic form that expands gradually in space 
(see Figure 3). This expansion is self-similar in time and can be well fitted
at large times by:
\be
n(x,t)=n_0(t)\left[1-\left(\frac{2x}{L(t)}\right)^2\right],\label{profile}
\ee
where $\pm \frac{L(t)}{2}$ are the points where the density vanishes. 
 We find,
for $N=50 000$,  $n_0(t) \simeq 725 \, t^{-0.3}$ and 
$L(t) \simeq 56.8 \, t^{0.35}$, compatible with $z=3$. This 
form suggests that $n_0(t)\propto 1/L(t)$, as expected for a scaling profile. 
The participation ratios indeed read in this case:
\be
Y_q=\frac{2^{2q}q!^2}{(2q+1)!} n_0^q (t) L(t) \propto t^{(1-q)/3}.
\ee
In particular, $Y_1=N=2 n_0(t) L(t)/3$, whereas the energy is expected to 
relax as $t^{-1/3}$,
a result that can be checked numerically independently. For such a parabolic 
profile,
the mean-squared radius is given by $R = L(t)/2\sqrt{5}$. 

Again, it is important to note that, except for the very last stages of the 
swelling,
the results are very similar whether the chain constraint is present or not; 
we will discuss this feature also in the following section. The final regime 
towards
the completely expanded configuration of the chain is hard to
study numerically as the chain gets trapped in some metastable states. The 
relaxation
towards the completely swollen state is slower and slower, as can be seen on 
Figure
1. 

\begin{figure}[htbp]
\begin{center}
\centerline{\psfig{figure=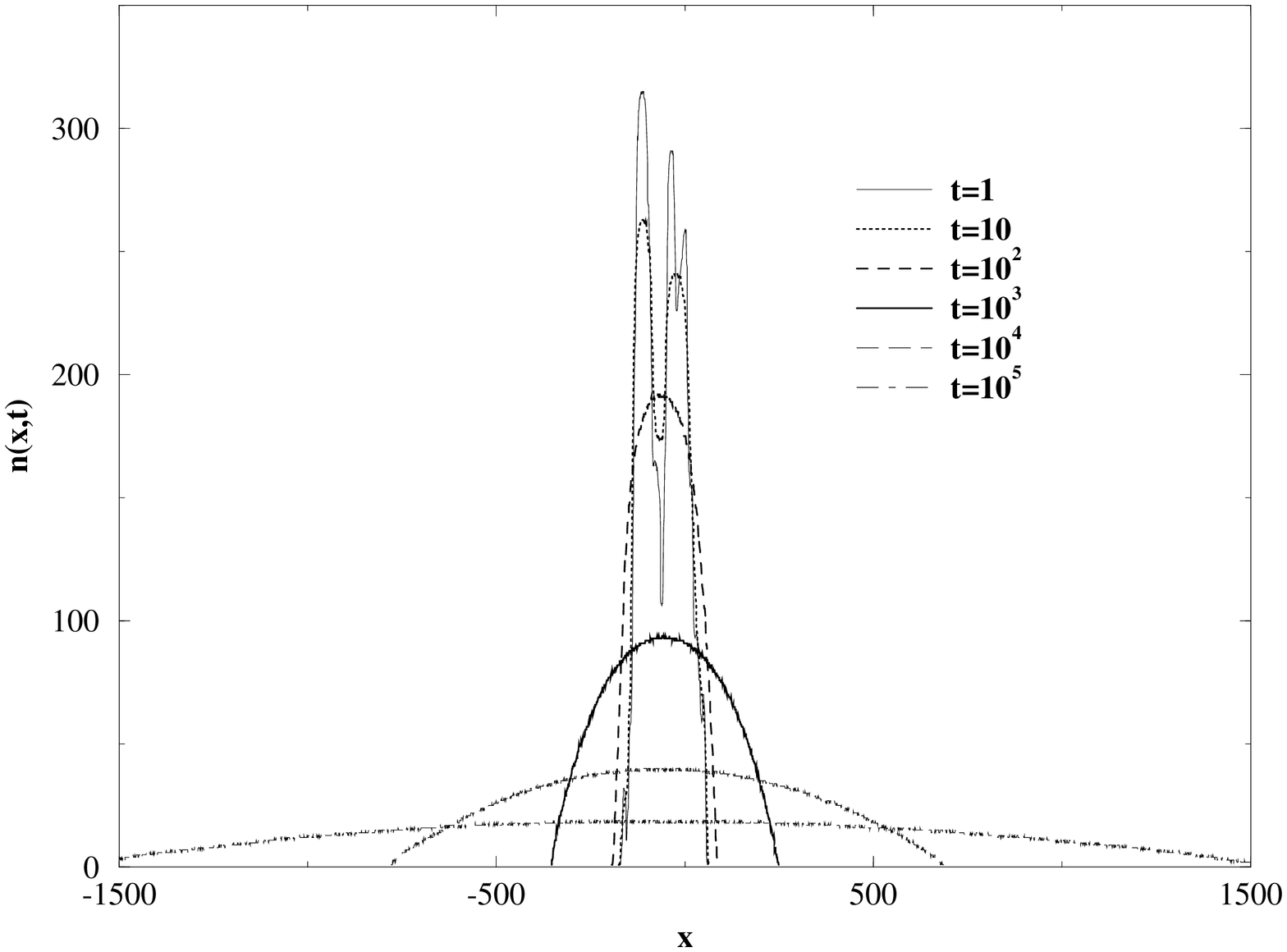,width=8cm,angle=270}}
\caption{Profiles $n(x,t)$ of the chain as it swells
from a random initial configuration, at times $t=1,10,10^2,...,10^5$, with 
$N=10000$.
The profile quickly becomes parabolic and expands in space as $t^{1/3}$.}
\end{center}
\end{figure}

In the next section, we present different analytical approaches to describe 
the swelling
mechanism to account for the above numerical results.

\subsection{Analytical theories for the swelling dynamics}

\subsubsection{A Flory theory}

A very crude analytical approach for the dynamics of the chain can be derived 
from a dissipative equation of motion for the size of the polymer $R$, 
combined with a Flory approximation for the free energy ${\cal F}(R)$.
The Flory free energy is as usual written as the sum of an entropic 
contribution
and an excluded volume interaction contribution. The resulting equation reads:
\be
\frac{\partial R}{\partial t}= -\mu_{chain} \frac{\partial {\cal 
F}(R)}{\partial R}=
-\frac{\mu_0}{N} \frac{\partial}{\partial R} \left[\frac{R^2}{N}+v 
\frac{N^2}{R^d}
\right],
\ee
where $\mu_{chain}$ is the mobility of the chain. It is well known 
\cite{deGennes} that 
the mobility for a chain is $\mu_{chain}=\frac{\mu_0}{N}$,
where $\mu_0$ is the mobility of a single monomer. The exact solution 
for this 
differential equation can actually be computed, and for an initial condition 
$R(t=0)=R_0$, 
reads:
\be
R(t)= e^{-\frac{2\mu_0 t}{N^2}} \left[R_0^{d+2} +
\frac{vd N^3}{2} (e^{\frac{2(d+2)\mu_0 t}{N^2}}-1)
\right]^{\frac{1}{d+2}}.
\ee
For large times, $R(t)$ converges exponentially fast towards the equilibrium 
Flory radius $R_F=({vd N^3}/{2})^{\frac{1}{d+2}}$:
\be
R(t) \simeq  R_F \left[1-\frac{1}{d+2}e^{-\frac{2 (d+2)\mu_0 
t}{N^2}}\right]
\ee
For an ideal chain initial condition, one has $R_0 \simeq \sqrt{N}$. 
Therefore, in the regime
where $N^{d/2}/v \ll \mu_0 t \ll N^2$, the initial condition term 
$R(0)^{d+2}$ can be neglected, and one finds:
\be
R(t) \simeq \left(\frac{d(d+2)}{2} v\mu_0 N t\right)^{\frac{1}{d+2}}.
\ee
Note that in this time regime, the elastic part in the Flory free energy (that 
comes from
the chain constraint) does not play any role and one therefore expects the 
same scaling
to be valid for an assembly of non-connected repulsive particles, in the 
intermediate
time regime mentioned above. As discussed above, this is indeed what we
find numerically, up to a rescaling of time.

In one dimension, the Flory theory predicts a power law behavior $R(t) \propto 
N^{\frac{1}{3}} 
t^{\frac{1}{3}}$ -- both scalings in rather good agreement with the numerical 
results.
The very short time behaviour predicted by this approach reads:
\be
R(t) \simeq R(0)\left(1+\frac{vd N\mu_0 t}{R(0)^{d+2}}\right)\label{Floryshort}
\ee
In one dimension, the departure from the initial condition is therefore
$R(t)=R(0)+\Gamma_F t^{\gamma_F}$
where $\gamma_F=1$ and $\Gamma_F$ is independent of the number of monomers 
$N$. This
approach therefore predicts that for short times $R(t)/\sqrt{N}$ is a function 
of
$t/\sqrt{N}$, in reasonable agreement with our numerical data (see Fig. 2).

A slightly more refined theory (but still in the spirit of Flory) 
actually allows one to understand the parabolic shape of the density profile. 
One can write a Master equation for the number of particles sitting on site 
$i$:
\be
\frac{\partial n_i}{\partial t}= \sum_{j} W_{ij} n_{j}(t) -\sum_{j} W_{ji} 
n_{i}(t)
\ee
where the transition probabilities at finite temperature $T=\frac{1}{\beta}$ 
are chosen to be:
\be
W_{ji}=W_{i \rightarrow j}= w_0 \Theta(n_j-n_i) 
\qquad \Theta(n_j-n_i)=
\frac{e^{\frac{v\beta}{2}[n_i-n_j-1]}}
{e^{\frac{v\beta}{2}[n_i-n_j-1]}+e^{-\frac{v\beta}{2}[n_i-n_j-1]}} 
\ee
In the zero temperature limit, the transition probabilities are simply 
described
by a step function:
\begin{eqnarray}\nonumber
W_{ji}=W_{i \rightarrow j}&=&  w_0  \quad {\rm if} \ \ n_i(t)>n_j(t)+1
\\
&=& 0  \quad {\rm if} \ \ n_i(t)<n_j(t)+1
\end{eqnarray}

In the simplest case where we allow moves to both nearest-neighbour sites,  
we require the condition that $W_{ij}=W_{ji}=0$ if $|i-j| \geq 2$. We neglect 
the chain condition, which is expected to be a valid approximation in the 
short
time limit. The Master equation then becomes, in one dimension ($d=1$):
\begin{eqnarray}\nonumber
\frac{\partial n_i}{\partial t}&=&  w_0  n_{i+1}(t) \Theta(n_{i}(t)-n_{i+1}(t))
+ w_0 n_{i-1}(t) \Theta(n_{i}(t)-n_{i-1}(t))
\\
&-& w_0  n_{i}(t) 
\left[\Theta(n_{i+1}(t)-n_{i}(t))+\Theta(n_{i-1}(t)-n_{i}(t))\right]
\end{eqnarray}
Let us assume that the density $n_i(t)$ is slowly varying in space, so that
one can expand the $\Theta$ function in powers of $\Delta n= n_{i+1}-n_i$ as:
$\Theta(\Delta n) \simeq \Theta_0 -\beta v \Delta n \Theta_1$. Then,
\be
\frac{\partial n_i}{\partial t} \simeq [n_{i+1}(t) + n_{i-1}(t) -2n_i(t)]+ g
[n_{i+1}^2(t) + n_{i-1}^2(t) -2n_i^2(t)],
\ee
where we have rescaled the time by $w_0 \Theta_0$, and set $g=\beta v
\Theta_1/\Theta_0$. Taking the continuous limit both in space, we end up with
the following non-linear conservative diffusion equation:
\be
\frac{\partial n(x,t)}{\partial t}= \left[\nabla^2 n + 2g \nabla (n 
\nabla n)\right].\label{nonlindiff1}
\ee

Let us note that the assumption $\beta v |\Delta n| \ll 1$ is crucial in the 
derivation of
this continuous equation. This may be acceptable in the case of repulsive 
interactions ($v>0$),
where the profiles $n(x,t)$ turn out to be smooth, so that the number
of monomers on neighbouring sites is almost the same. Surprisingly, however, 
this non
linear equation
describes also very well the zero temperature results in the long time limit. 
This is not true 
in the case of attractive interactions ($v<0$): the above equation then 
becomes unstable. This correspond to a `pearling instability' in the collapse 
regime
which we will discuss further in Section 4.

The non-linear diffusion equation (\ref{nonlindiff1}) admits a self-similar 
solution for the
profile $n(x,t)$, of the form:
\be
n(x,t)=\frac{N}{R_s(t)} f\left(\frac{x}{R_s(t)}\right)
\ee
where $R_s(t) \equiv t^{\alpha}N^{\delta}$ is the scaling form for the 
end-to-end distance of the chain in the intermediate time regime that we 
study, and $f$ a scaling
function independent of both $t$ and $N$. Introducing the rescaled variable 
$u={x}/{R_s(t)}$, we find the following differential equation for $u$:
\be
-\frac{\alpha N}{N^{\delta}t^{\alpha +1}} [f(u)+u f'(u)]=
 \frac{N}{t^{3 \alpha}N^{3 \delta}} f''(u)
+2g \frac{N^2}{t^{4 \alpha} N^{4 \delta}} [f'^2 (u) +f(u) f''(u)]
\ee
We assume that in the regime of time studied here, the diffusion term
can be neglected, which is true if $t^{\alpha} \ll g N^{1-\delta}$,
and can be checked {\it a posteriori} when $R_s \ll N$. Then the equation 
reduces to:
\be
-\alpha N^{1- \delta} t^{3 \alpha -1} [f(u)+uf'(u)]=
2 g  N^{2 -4 \delta} [f'^2 (u) +f(u) f''(u)]
\ee
This fixes the values of both exponents: $\alpha_F=\frac{1}{3}$
and $\delta_F=\frac{1}{3}$, which gives back the results of the Flory equation
of motion for $d=1$. Furthermore, the resulting ordinary differential
equation for $f(u)$ has a parabolic solution. The final result, normalized to
give the correct number of monomers, is Eq. (\ref{profile}), with 
$n_0(t)=N^{2/3}t^{1/3} (3/64 g)^{1/3}$ and
$L(t)=2 (9\, g\, N\, t)^{1/3}$, in agreement with the numerical results of 
Fig. 3 (obtained at zero temperature).

\subsubsection{A dynamical variational method}

An alternative, approximate method to study the dynamics of swelling and 
collapse of
polymer chains was proposed in \cite{Orland}. It is based on a dynamical 
extension of the
effective Kuhn length method of Edwards, which is known to reproduce the 
Flory value for the end-to-end distance exponent. The essence of the method, 
together with the
self-consistent dynamical equations, are recalled in Appendix A. From these 
equations, the
short time and long time dynamics of the chain was obtained in \cite{Orland}. 
In
the short
time limit, one finds that the chain radius grows, in one dimension, as:
\be
R(t)=R_0 + \Gamma_v t^{\gamma_v} \qquad \gamma_v=\frac{5}{4}
 \qquad \Gamma_v \propto N^0,
\ee
at variance both with the Flory prediction $\gamma_F=1$, and with the 
numerical simulations. 

The intermediate self-similar expansion regime was not worked out in 
\cite{Orland}. Looking for a self similar solution of the form 
$R(t) \propto t^{\alpha}N^{\delta}$, 
we have found (after some work) the following values for $\alpha$ and $\delta$:
\be
\alpha_v=\frac{1}{14} \qquad \delta_v=\frac{11}{14},
\ee
Quite surprisingly, again, these values differ from the simple Flory 
prediction and are very
far from the numerical result. From the above value of the exponents, one
predicts a stretching time of the order of $N^3$ (longer than what is observed 
numerically, where the stretched regime is obtained when $\tau=t/N^2$
becomes of order $1$). 

It is intriguing to see that a somewhat more refined theory
(but which basically contains also the ingredients of a Flory theory) gives 
worse
results than the simple-minded approach. It seems actually that the Gaussian 
variational
method misses the fact that the dynamics is governed by repulsive interactions 
but
not sensitive (at short times) to the chain constraint. Hence, the `effective' 
chain
approximation is very bad in this region. The approximation might be better at 
{\it long} times, to describe the approach to the swollen state when the
chain constraint becomes relevant. We have shown in Fig. 1 the predicted 
$t^{1/14}$ for comparison; however, as discussed in the next paragraph,
a logarithmic growth is probably more adapted to describe this regime.

\subsection{Late stages of the dynamics}

It can be observed from the Monte Carlo simulations (see Figure 1) that the 
dynamical evolution becomes frozen at $T=0$, while it is strongly slowed down  
at finite temperatures. In Figure 4, we show snapshots of the configuration
of a short chain, with $N=100$, as it evolves with time. 

\begin{figure}[htbp]
\begin{center}
\centerline{\psfig{figure=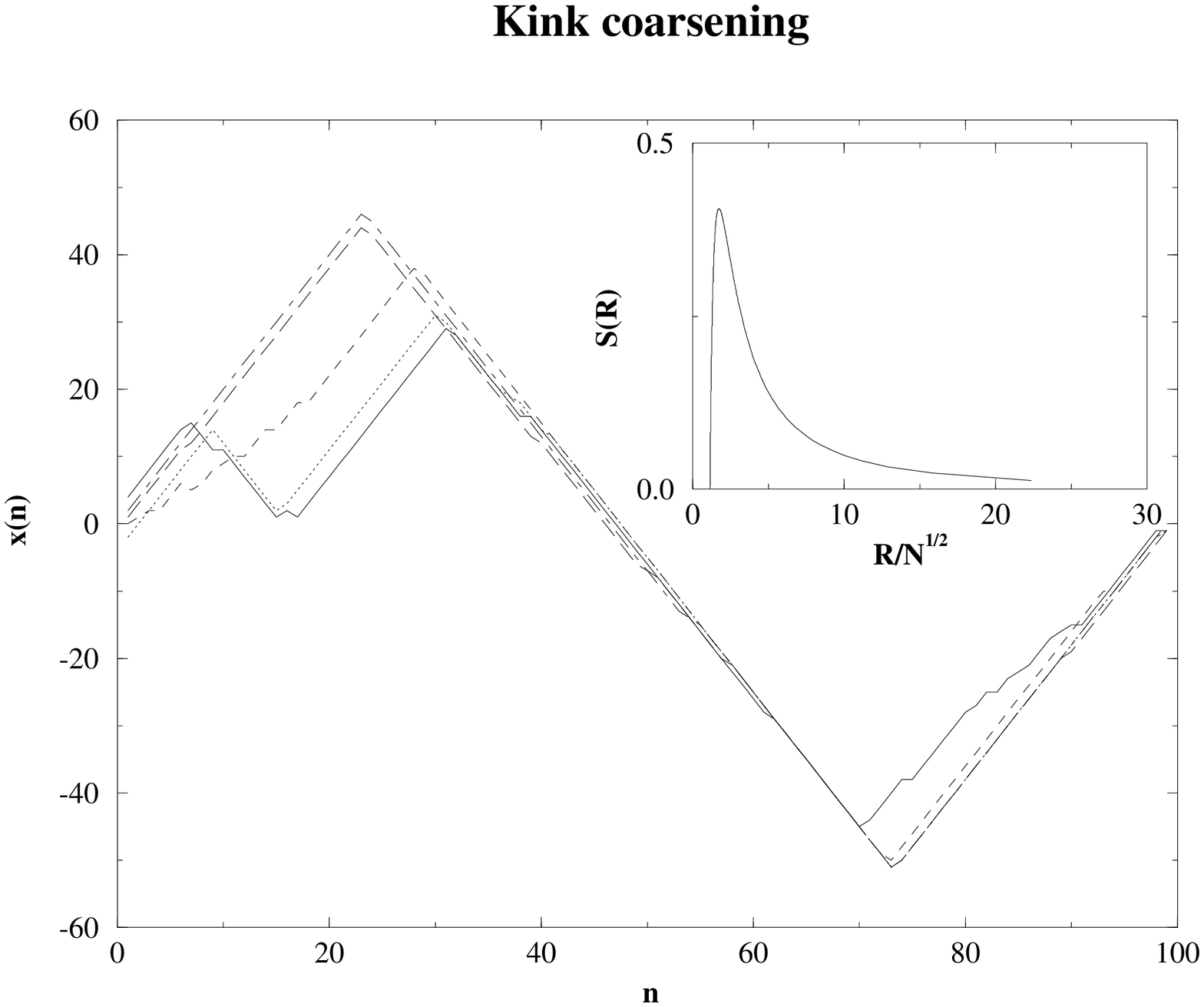,width=8cm,angle=270}}
\caption{Snapshots of the configuration
of a short chain, with $N=100$, as it evolves with time. The $x$ axis is 
the internal coordinate of the monomers, while the $y$ axis is the 
corresponding position in space. One can see that the long time dynamics 
consists of
the progressive disappearance of `kinks'. Inset: The complexity ${\cal S}$ 
as a function of $R$.}
\end{center}
\end{figure}

It is clear that
the chain is trapped for long period of times in metastable configurations,
consisting of fully stretched segments of the chain going alternatively
to the left or to the right. If we call $2M$ the total number of segments and
$\ell$ their average size, the typical extension of the chain in such a
configuration is $R \sim \ell \sqrt{M}$. Since $N=2M\ell$, one has:
$R \sim N/\sqrt{M}$, or $M \sim r^{-2}$. The dynamics of the chain in the
long time region therefore consists of a slow `coarsening', where `kinks'
progressively disappear, leading to a decrease of $M$. The time needed 
for a hairpin of size $\ell$ to disentangle can be estimated from the
energy barrier to be crossed, which is proportional to $\ell$ itself.
Therefore, one has $\ln t(\ell) \sim \beta v \ell \sim \beta v N r^2$.
From this argument, we obtain a logarithmic growth law of the size of the
chain in the long time regime:
\be
R \sim \sqrt{\frac{N}{\beta v} \ln t},
\ee
in qualitative agreement with the numerical simulations: We show on Fig. 5
a logarithmic fit of the radius of gyration as the last stages 
of the stretching occur. The above estimate assumes that all segments
lengths are close to the average value. This is confirmed by an analytical 
calculation of the number of metastable states of the stretching
chain, and is explained in detail in Appendix B.

\begin{figure}[htbp]
\begin{center}
\centerline{\psfig{figure=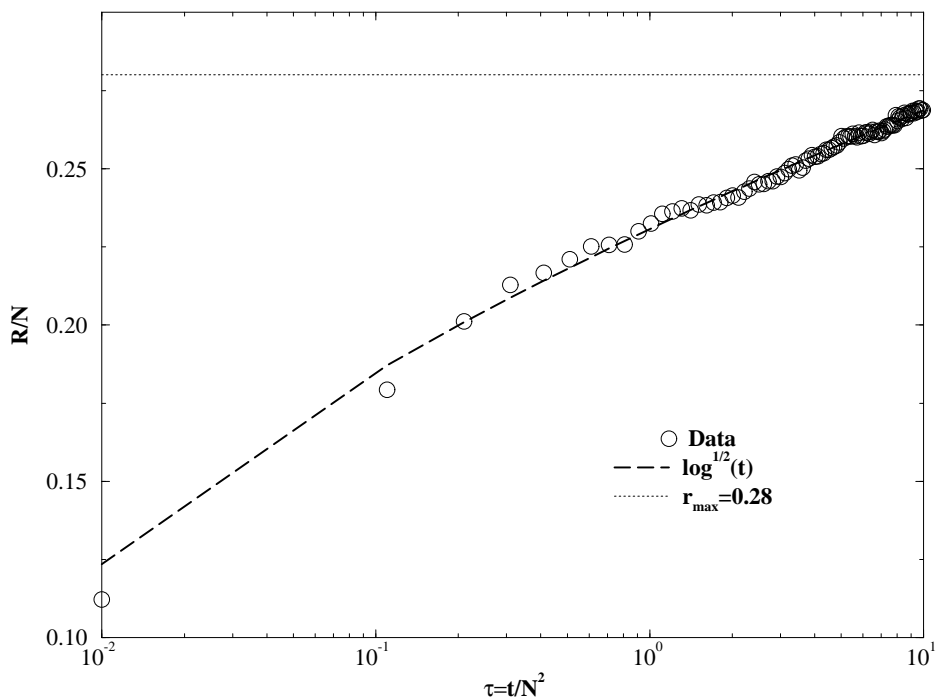,width=8cm,angle=270}}
\caption{Logarithmic growth of the chain at the end of the swelling
dynamics, for a small chain $N=100$}. \end{center}
\end{figure}

It is indeed
interesting to count the number ${\cal N}(M)$ of metastable states
of the chain for a fixed value of $M$, i.e., for a given size $R$ of the chain.
The problem is actually related to counting the metastable states of the 
one dimensional Ising model with Kawasaki dynamics; the corresponding 
calculations are given in Appendix B. 
The idea is to make an analogy between a segment of the chain going to the
right (or to the left)
and a domain of plus (or minus) spins. An additional 
constraint is needed to make the configuration metastable. The chain can only
evolve by moving inwards the monomer at the extremity of the hairpin (see Fig. 
4).
This is possible if the total number of monomers sitting on the target site is
less than the number of monomers at the starting site. If we assume that the 
local
density of monomers is random, the probability for this to happen is equal to 
$1/2$.
Therefore, a zig-zag configuration with $M$ segments is stable with 
probability $2^{-M}$.
The result is that ${\cal N}(M) \sim
\exp(N{\cal S}(M))$, with a complexity ${\cal S}$ given by: 
\be {\cal
S}=-\delta \ln 2 +(\delta-1) \ln(1-\delta)  -\delta
\ln \delta, \ee 
with $\delta=2M/N \sim N/R^2$. This
function is plotted in the inset of Figure 4. At the
beginning of the swelling, the complexity increases very rapidly with $R$ ; 
after a maximum reached for $\delta^*=1/3$, ${\cal S}$ then decreases
at large $R$.

If one adds all frozen configurations for all possible values of $M$ (or
equivalently, $R$), one finds a total number of frozen states which grows as
${C^N}/{\sqrt{N}}$ where the complexity per monomer is $C=3/2$.
This must be compared to the total number of configurations, which is equal,
for $m=1$, to the number of closed random walk, i.e. ${2^N}/{\sqrt{N}}$.

\section{Attractive case: collapse of the chain}

\subsection{Numerical Results}

The case of attractive interactions can be studied numerically
in the same way as before, except that now the interaction parameter 
$v$ is negative. In this case, obviously, the ground state of the chain 
is such that all monomers are located on the same site. As the chain 
collapses, the
profile $n(x,t)$ very
quickly becomes discontinuous as monomers aggregate in well separated 
``clusters'':
see Figure 6. 

\begin{figure}[htbp]
\begin{center}
\centerline{\psfig{figure=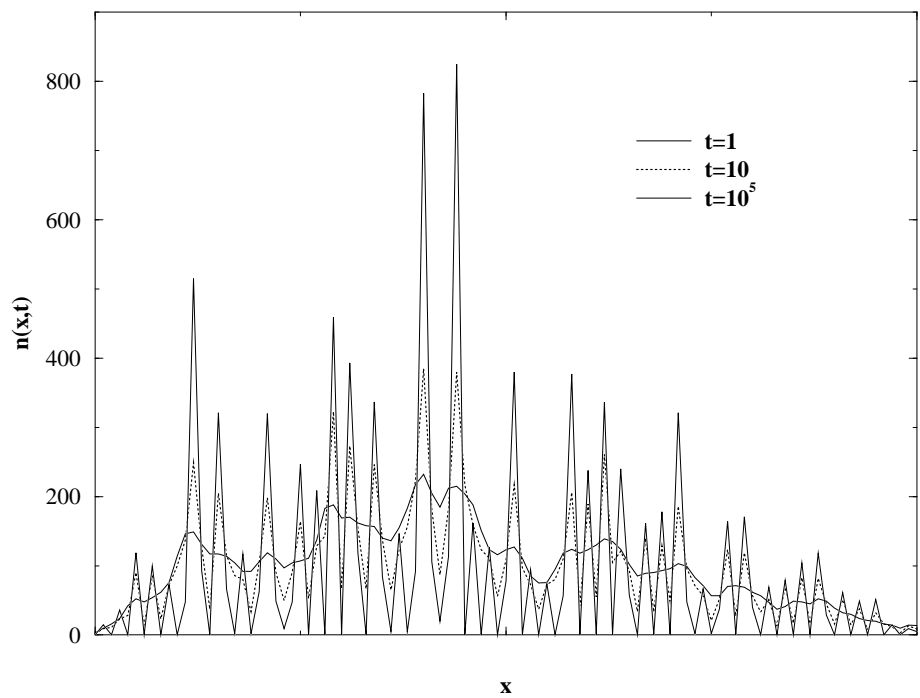,width=8cm,angle=270}}
\caption{Profiles $n(x,t)$ of the chain as it collapses
from a random initial configuration, for $T=5$, at times
$t=1,10,10^5$. The profiles obtained for $t=10^2, 10^3$ and $10^4$ 
are indistinguishable to the eye from the last one.}
\end{center}
\end{figure}

These clusters are very slow to coalesce and the dynamics is slower and slower 
with time
(and as the temperature $T$ decreases).
This slowing down is due to the fact that
it is energetically very unfavorable to `break' one of these
clusters in order to transfer its monomers to larger clusters.
The energy needed to remove one monomer from a cluster containing $n$
monomers is equal to $n|v|$ (for $n$ large), corresponding to a time
$\propto \exp(\beta|v| n)$. Therefore, the system remains trapped longer and 
longer in
configurations where there are $M$ clusters containing $n \sim N/M$
particles. This can be seen for example on the time evolution of the
energy $E(t)$: single runs for three different temperatures are shown 
in Figure 7. By averaging $E(t)$ over several {\sc mc} runs, we found a 
logarithmic decay
with time, compatible with the estimate $E \sim - M n^2 \sim -N \ln(t)$, using 
$n \sim T\ln t$. It is interesting to note that the dynamics of clusters is 
very close
to that of the one dimensional `Backgammon' model introduced \cite{Ritort} and 
further
studied in \cite{GBM,GodLuck,Ritortbis}: although the barriers are `entropic' 
in the
latter model,
the average time needed to empty a cluster also grows exponentially with its 
size $n$.

\begin{figure}[htbp]
\begin{center}
\centerline{\psfig{figure=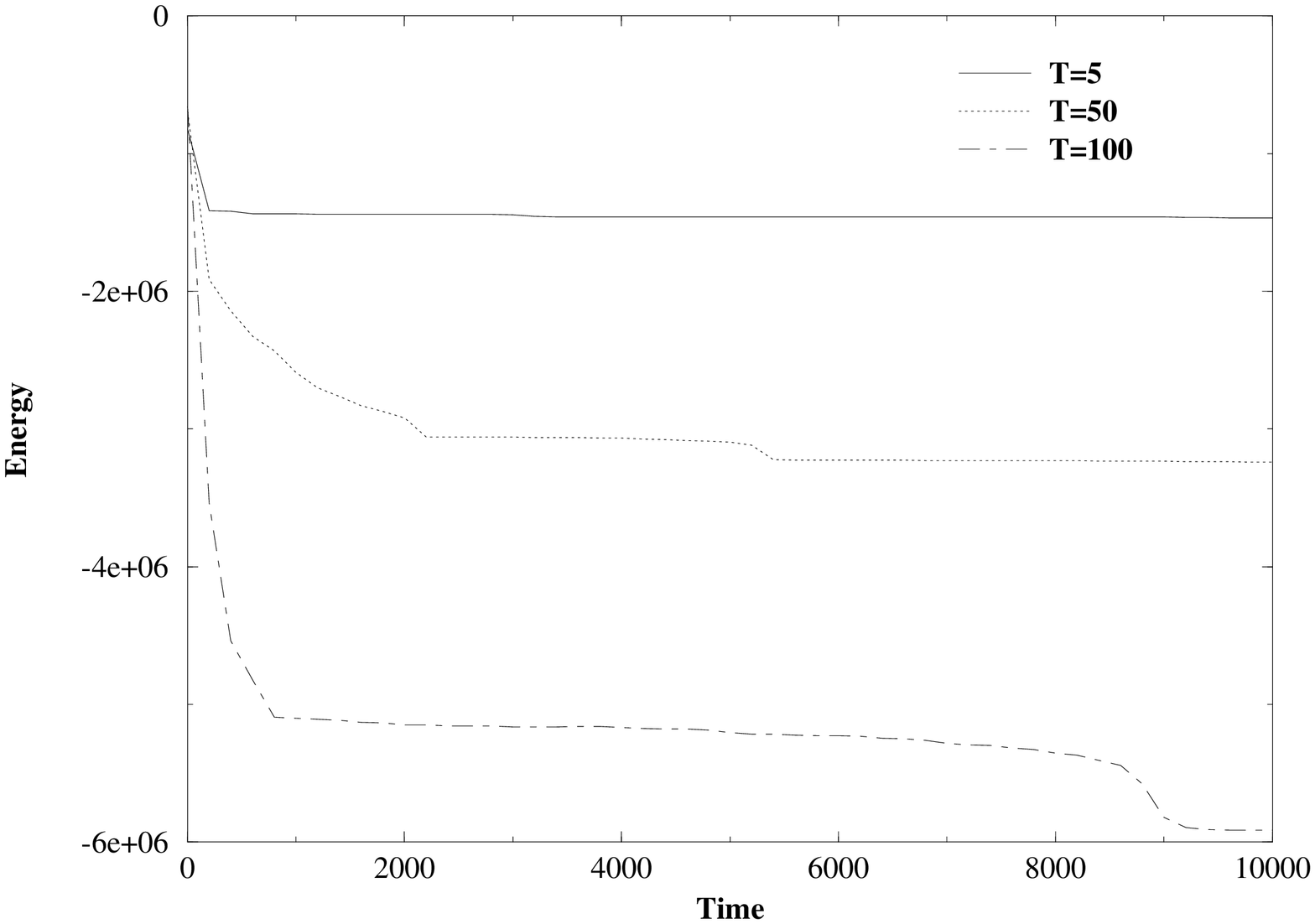,width=8cm,angle=270}}
\caption{Energy of the polymer as a function of time
for a single Monte-Carlo run, at $T=5,50,100$:
energetic traps can be visualized.}
\end{center}
\end{figure}

\subsection{Breakdown of the continuum description}

Let us come back to the continuous approach developed
in the previous section in the case of repulsive interactions.
Starting again from the master equation, and considering attractive
interactions this time, we find the following
equation for $n(x,t)$:
\be
\frac{\partial n(x,t)}{\partial t}= \left[\nabla^2 n - 2|g| \nabla (n 
\nabla n)\right].\label{nonlindiffbis}
\ee
A naive extension of the self-similar solution found in the repulsive
case leads to the following equation for the size of the polymer $R_s$:
\be
\frac{dR_s(t)}{dt}=-2 |g| \frac{N}{R_s^2(t)},
\ee
or $R_s^3(t)=R^3(0) -6N|g|t$, suggesting that all monomers coalesce 
into a single cluster in a finite time. This clearly does not capture
the essential physical feature of the dynamics at late time, namely
the activated coarsening of well separated `towers' of monomers.

In fact, Eq. (\ref{nonlindiffbis}) leads to a short wavelength 
`pearling' instability of initially smooth profiles. Starting from 
$n(x,t)=n_0+\delta n(x,t)$,
with $\delta n \ll n_0$, one finds: 
\be
\frac{\partial \delta n(x,t)}{\partial t}= (1-2|g|n_0) \nabla^2 \delta n, 
\label{nonlindiffter}
\ee
which is unstable as soon as $2|g|n_0 > 1$, i.e. for small enough 
temperatures or when the chain density is large. As observed numerically, 
a random walk initial
condition is indeed immediately unstable for large $N$, since $n_0 \sim 
\sqrt{N} \gg 1$.

As soon as the chain has entered the pearled state with well separated 
clusters of monomers, any continuous description is bound to fail. A 
discussion in terms of metastable states, and jumps between these states,
is needed. 

\subsection{Calculation of the number of metastable states}

A metastable state of lifetime $t$ is a configuration of clusters 
which remains frozen for a time $t$. This means that (i) two clusters
cannot be nearest neighbours and (ii) the minimum 
number of monomers in each of the clusters at temperature $T$ 
is at least equal to $k^*(t,T)=\max(2,T\ln(t))$ -- since a cluster
must contain at least two monomers. How many such configurations are
there ? This is a combinatorial problem including two steps: first 
count the number of configurations with $N$ monomers and $M$ clusters each 
containing
more than $k^*$ monomers, then the number of arrangements of $M$ clusters on a 
one-dimensional box of $L$ sites, so that all clusters are at least separated 
by
an empty site. We will restrict to the case where the
chain constraint is $m=1$, i.e., two consecutive monomers cannot be 
at a distance larger than $1$. In this case, the number of monomers 
within a `hole' is exactly equal to $2$, because the chain is closed
onto itself. Therefore, the total number of monomers belonging to clusters
is related to the length $L$ occupied by the chain by:
\be
\sum_{i=1}^{M} n_i = N-2(L-M).
\ee
The number of configurations 
${\cal N}_{k^*} (M|N)$ with at least $k^*$ monomers
in each cluster can be calculated using the following formula:
\be
{\cal N}_{k^*} (M|N) = \sum_{n_1=k^*}^{N} \dots \sum_{n_M=k^*}^{N} 
\delta\left(\sum_{i=1}^{M}
(n_i-2) -N +2L\right),
\ee
which can be transformed into:
\be
{\cal N}_{k^*} (M|N) = \int d\lambda e^{-i \lambda (N-2L)} \left(
\sum_{n=k^*}^{N} e^{i\lambda (n-2)}\right) ^M=\int d\lambda e^{-i \lambda N} 
\left(
\frac{e^{i(k^*-2)\lambda}}{1-e^{i\lambda}}\right)^M
\ee
We define in the following $\delta=\frac{M}{N}$, $\rho=N/L$ and $z=i\lambda$, 
and perform
a saddle point approximation on the above integrand in the large $N$ limit. We 
find that
at the saddle point,
 $z^*=\ln\left(\frac{1-\frac{2}{\rho}-\delta (k^*-2)}
{1-\frac{2}{\rho}-\delta (k^*-3)}\right)$.
Then, we obtain:
\be
\ln\left[{\cal N}_{k^*} (M|N)\right]=
\left[-1+\frac{2}{\rho}+\delta (k^*-2)\right] \ln\left(\frac{1-\frac{2}{\rho}-\delta
(k^*-2)}
{1-\frac{2}{\rho}-\delta (k^*-3)}
\right)
-\delta  \ln\left(\frac{\delta}{1-\frac{2}{\rho}-\delta (k^*-3)}\right).
\ee

Now we count the number of ways ${\cal N}_L(M)$ to arrange $M$ clusters
on $L$ sites with all the clusters at least separated by an empty site.
This can be done in the following manner, mapping this problem into an 
antiferromagnetic spin representation. (A more direct combinatorial method
can be found in \cite{Ritortbis}). We define the formal antiferromagnetic 
hamiltonian:
\be
{\cal H}= J \sum_{i=1}^L \sigma_i \sigma_{i+1}, \ \ J>0,
\ee
where $\sigma_i=1$ by definition if there is at least one monomer on site $i$,
and $\sigma_i=0$ if site $i$ is empty, so that configurations with consecutive 
occupied
sites have a positive energy and the configurations
of zero energy are those with at least one empty site
between each occupied site. Using the additional constraint that the total
number of occupied sites is set to some value $M$, the number of configurations
${\cal N}_L(M)$ is given by the following formula:
\be
{\cal N}_L(M)=\lim_{J \to \infty} \sum_{\{\sigma_i=0,1\}} e^{-{\cal H}}
\delta (M-\sum_{i=1}^L \sigma_i).
\ee
This can be rewritten using the integral representation
of the $\delta$-function:
\be
{\cal N}_L(M)= \int d\mu e^{i \mu M} Z_L(i\mu)
\ee
where $Z_L(i\mu)= \sum_{\{\sigma_i=0,1\}} e^{- {\cal H}}
e^{-i \mu \sum_{i=1}^L \sigma_i}$.
By performing a saddle-point approximation on $\mu$,
one has to find $\mu^*$ such that
\be
M + \frac{\partial}{\partial(i \mu)}|_{\mu=\mu^*} \ln Z_L(i \mu)=0.
\ee
The quantity $Z_L(i \mu)$ can actually be computed exactly using a 
transfer-matrix method,
and in the limit $J \to \infty$,
$Z_L(i \mu)=\left[\frac{1}{2} (1+ \sqrt{1+4e^{-i \mu}}\right]^L$.
Using the saddle-point equation, $\mu^*$ is easily found to be equal to
$\sqrt{1+4e^{-i \mu}}=\frac{1}{1-2\delta\rho}$. Finally,
\be
{\cal N}_L(M)\simeq  e^{L\Sigma(\delta\rho)},
\ee
with
\be
\Sigma(\delta\rho)=
i\mu^* \frac{M}{L} + \ln Z_L(i\mu^*)=
-\delta\rho \ln\left(\frac{\delta\rho(1-\delta\rho)}{(1-2\delta
\rho)^2}\right)
+\ln\left(\frac{1-\delta\rho}{1-2\delta\rho}\right).
\ee
One can check in particular that one finds $\Sigma(0)=0$ and 
$\Sigma(\frac{1}{2})=0$,
as it should.

We now combine the two above results, and deduce the total number of 
$k^*$-stable
configurations for a given density $\rho$:
\be
{\cal N}_{\rho} (k^*)\equiv e^{N S_{\rho}(k^*)}=\sum_{M=1}^N {\cal N}_L(M)
{\cal N}_{k^*} (M|N).
\ee
This is again computed from a saddle point (in $M$) and leads to: 
\bea\nonumber
S_{\rho}(k^*)&=&(-1+\frac{2}{\rho}+\delta (k^*-2)) 
\ln\left(\frac{1-\frac{2}{\rho}-\delta (k^*-2)}
{1-\frac{2}{\rho}-\delta (k^*-3)}
\right)
-\delta  \ln\left(\frac{\delta}{1-\frac{2}{\rho}-\delta (k^*-3)}\right)\\
&-&\delta^* \ln\left(\frac{\delta^*\rho(1-\delta^* \rho)}{(1-2\delta^* 
\rho)^2}\right)
+\frac{1}{\rho} \ln\left(\frac{1-\delta^* \rho}{1-2\delta^* \rho}\right)
\eea
In this expression, $\delta^*$ is the saddle point value of $\delta$, defined 
as:
\be
\frac{(1-\frac{2}{\rho}-\delta^*(k^*-2))^{k^*-2}
(1-\frac{2}{\rho}-\delta^*(k^*-3))^{3-k^*}}{\delta^*}
\frac{(1-2\delta^*\rho)^2}{\delta^*\rho(1-\delta^*\rho)} =1
\ee

At zero temperature, all metastable states are frozen, and the size of
the chain $L$ and therefore the density $\rho$, are constant in time. One
can estimate the complexity by setting $k^*=2$ in the previous formulae.
We found numerically the solutions of the above equations $\delta^*(\rho)$
and $S(\rho)$ and plotted the results in Figure 8. One should note that since 
each
cluster contains at least 2 monomers, the maximum number  $M$ of clusters is 
$\frac{N}{2}$, so $\delta \leq \frac{1}{2}$. Since a metastable state 
is such that two clusters should be separated by at least
one empty site, one also has $L \geq 2M$, or $\delta \rho \equiv \frac{M}{L} 
\leq \frac{1}{2}$. 

\begin{figure}[htbp]
\begin{center}
\centerline{\psfig{figure=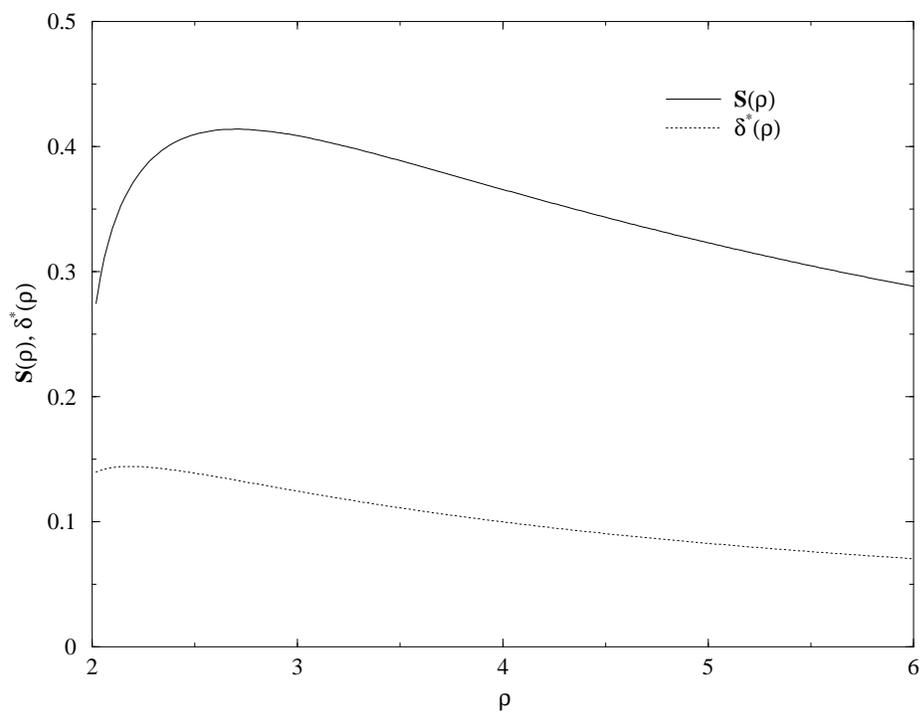,width=8cm,angle=270}}
\caption{Entropy of metastable states $S(\rho)$ and the corresponding optimal 
number
of clusters $\delta^*(\rho)$, as a function of 
the density of monomers $\rho$, at $T=0$.}
\end{center}
\end{figure}

It is interesting to draw an analogy with some recent work on the
number of metastable states in spin-glasses \cite{Biroli}. As $k^*$ increases,
one counts  configurations which are more and more stable. Similarly, in 
spin-glasses,
one can introduce $k^*$-spin flip stable configurations, i.e. configurations
the energy of which cannot be lowered by any flip involving at most $k^*$ 
spins.
The complexity of these states was determined for simple spin-glasses by  
Biroli and
Monasson \cite{Biroli}. They show
that in mean-field models, inherent states (corresponding to 1-spin flip 
stable states),
pure states and $k^*$ stable states all coincide.
But they also show that in finite dimensions (contrarily to infinite 
connectivity models)
all these different concepts are distinct and actually help to visualize the 
degree of
freezing of a glassy system, according to the lifetime of the metastable 
states relevant
after a certain time scale. In
the following section, we will show that the analysis of the relevant 
metastable
states of a collapsing polymer allows one to predict how the radius of
gyration of the chain decreases with time.

\subsection{Consequences for the evolution of the chain at non zero
temperatures} 

At non zero temperatures, the relevant metastable states after a time
$t$ are such that $k^* \sim T \ln t$. Furthermore, it is reasonable to 
assume that the chain size is such that the number of metastable states
is maximal, that is to say, the chain is in one of 
its most probable configurations 
after a time $t$. We thus maximize the entropy
and obtain $\rho(t)=\rho^*(t)$ (and
therefore  $L(t)=N/\rho^*(t)$) 
by setting:
\be
\left.\frac{dS_{\rho}(k^*)}{d\rho}\right|_{\rho=\rho^*}=0,
\ee

Solving for $\rho^*$ and $\delta^*$ in the large $u=T\ln t$ limit, we
find:
\be
\rho^*(u) \simeq 4 \ln u \qquad \delta^*(u) \simeq \frac{1}{u} -\frac{1}{u\ln 
u},
\ee
and the corresponding value of the entropy, to leading order in $u$:
\be
S^*(u) \simeq 2 \frac{\ln u}{u} 
-2\frac{\ln(\ln u)}{u} +\frac{1}{u}(4-\ln 8).
\ee

These results predict a slow decrease of the number of clusters and an 
extremely slow
decrease of the size of the chain:
\be
M(t) \sim \frac{1}{T\ln t} \qquad L(t) \sim \frac{1}{4 \ln(T\ln t)} \qquad
S(t) \sim \frac{2\ln(T\ln t)}{T\ln t}.
\ee
This last result is quite interesting and is a non trivial consequence of the
present calculation of the complexity of metastable states. It is obviously
very hard to check numerically, but suggests an hyper-slow contraction of 
the chain. Numerically, we find that the chain length indeed hardly decreases
with time.

\begin{figure}[htbp]
\begin{center}
\centerline{\psfig{figure=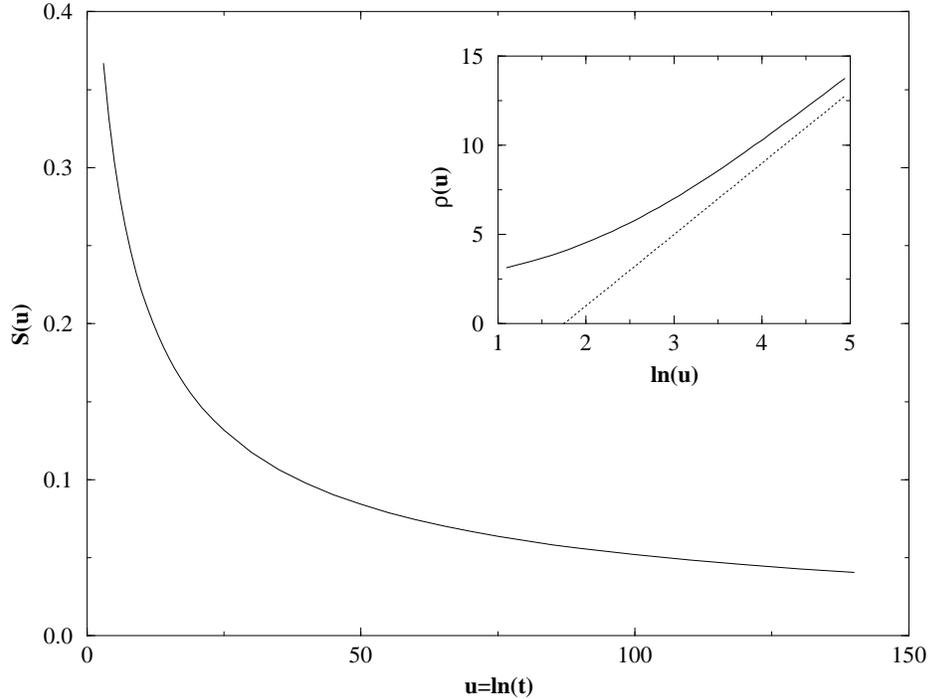,width=8cm,angle=270}}
\caption{Entropy of metastable states $S^*(u)$ as a function
of $u=T\ln(t)$. Inset: Density of the chain as a function
of $\ln(u)$ (plain line). The dotted line is the asymptotic result 
$\rho(u)\simeq 4\ln(u)$.}
\end{center}
\end{figure}

\subsection{Calculation of the density of metastable states at fixed 
energy}

We now turn to the calculation of the number of metastable states
corresponding to a certain degree of freezing $k^*$, and to a fixed total 
energy
of the system $E$. As was done before in the case of spin-glasses
\cite{TAP}
or Josephson arrays \cite{Josephson}, 
we want to understand how the entropy of metastable states
behaves as a function of energy at $T=0$, and we go here
further by looking at its dynamical
evolution with respect to $k^*=T \ln t$.
The calculation proceeds much in the same way as in the previous section
(we also restrict ourselves to the case where $m=1$),
starting from the expression of the 
number of $k^*$-stable configurations with $M$ clusters
and $N$ monomers, at fixed energy $E$:
\be
{\cal N}_{\rho,k^*} (E,M|N) = \sum_{n_1=k^*}^{N}
\dots \sum_{n_M=k^*}^{N} \delta\left(\sum_{i=1}^{M} (n_i-2) -N+2L\right)
\delta\left(\sum_{i=1}^{M} -\frac{1}{2} n_i(n_i-1)-(L-M)-E\right).
\ee
The technical details of the calculation are relegated to Appendix C.
Setting $\epsilon=\frac{E}{N}<0$ and $S_{\rho}(\epsilon,k^*)$ the complexity
per monomer, our final result is plotted on Figure 10. 
We see
that all $S_{\rho}(\epsilon, k^*)$ for different $k^*$ follow a common envelope
for a given $\rho$,
and asymptotically the curves approach the following law
at very low energies:
\be
S_{\rho}(\epsilon) \simeq_{\epsilon \to -\infty}
-\frac{\ln(-\epsilon)}{\epsilon} (1-\frac{2}{\rho})^2
+\frac{1}{\epsilon}(1-\frac{2}{\rho})^2 [-1+\frac{1}{2}
\ln(\frac{\rho}{4}(1-\frac{2}{\rho})^{3/2})] + o(\frac{1}{\epsilon^2})
\ee
For large times, $\rho \to \infty$ and we are left with:
\be
S_(\epsilon) \simeq_{\epsilon \to -\infty} -\frac{\ln(-\epsilon)}{\epsilon}
\ee
Even for very low energies, the entropy of metastable states never reaches 
zero:
contrarily to the example of spin-glasses, there is no lower cut-off in the 
energy
per
monomer below which the number of metastable states is not exponentially large 
in
$N$.

\begin{figure}[htbp]
\begin{center}
\centerline{\psfig{figure=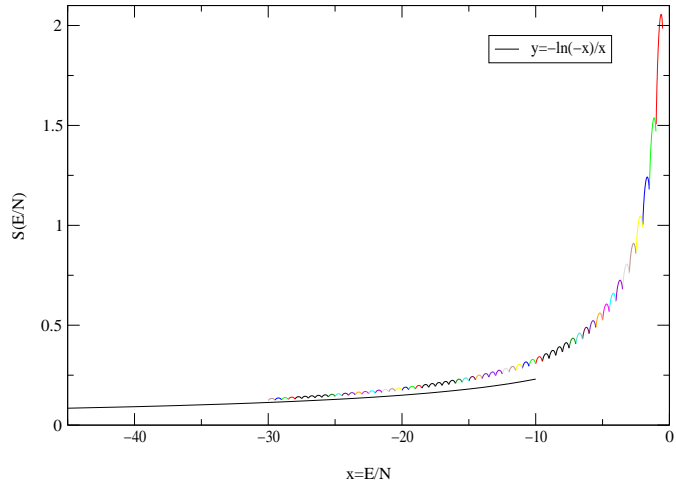,width=8cm}}
\caption{Entropy of metastable states $S_{\rho}(\epsilon)$ as a function of
the energy per monomer.
 The plain line is the asymptotic law  $S(\epsilon)=-\ln(-\epsilon)/\epsilon$.
For a given $k^*$, the domain of definition of the entropy is bounded 
by $\epsilon_{max}(k^*)=-\frac{1}{2}(k^*+1)+\frac{k^*}{\rho}$.
The successive arches, defined on $[\epsilon_{max}(k^*+1),
\epsilon_{max}(k^*)]$, correspond from right to left, to increasing $k^*$
($=3,4,..$),  and follow a common envelope.}
\end{center}
\end{figure}

\subsection{Distribution of cluster sizes}

Let us define $p(n,t)$ as the distribution 
of sizes of the clusters, at a given time $t$ and temperature, $n$ being the 
number of monomers in a given metastable cluster. We further define $\overline 
n$
as the average number of monomers in a cluster. From the above sections, we
know that $\overline n(t)$ grows slowly (logarithmically) with time. 
We have studied $\overline n p(n,t)$ as a function of $z=n/\overline n$ at different times. These different curves collapse well onto a single one
(note however that the range of variation of $\overline n$ cannot be very large). 
The scaling function $\tilde p$ is very far from a simple exponential, as
a naive maximum entropy argument would suggest. In particular, the tail 
of $\tilde p$ can be fitted as a power-law $\tilde p(z)\sim (z+z_0)^{-1-\nu}$,
with $\nu \simeq 1.9$. For $z$ small, on the other hand, $\tilde p(z)$ 
rapidly vanishes. Although our statistics is not very good, one can fit 
$\ln \tilde p(z)$ as $(z_1/z)^c$, with $c \sim 1$. 

The present cluster dynamics is actually quite similar to the 
so-called `cut-in-two' model introduced by Derrida et al.
\cite{Godreche} to describe the pattern of coalescing
droplets in one dimension. In the latter model, a collection
of intervals of different sizes (which represent
the droplets) evolves according to the following simple rule: one picks
the smallest interval, divides it in two equal parts and sticks the right part
to the right neighbour, while the left part coalesces with the left 
neighbour.
In this model, the average size of the intervals grows with time, and
the rescaled distribution of interval sizes tends to an asymptotic
distribution shown in Fig. 11. The cluster dynamics studied here is
similar for the following reason: because the time needed to empty one
cluster is exponential in its size, the first cluster to disappear will 
typically be the smallest one available after time $t$. If one
neglects the chain constraint, the smallest cluster empties itself 
by sending an equal number $n/2$ (on average) of monomers towards its two
immediate neighbours. The analogy is not exact, though, since this
number is not exactly equal to $n/2$ at each iteration; furthermore, some
subtle correlations are induced by the chain constraint. One can nevertheless
take the simple `cut-in-two' model, or a randomized version where a random
fraction $f$ is pasted to the left and $1-f$ to the right, as benchmarks to
which we can compare our results. This is performed in Fig. 10. The agreement 
is only fair, and is better for the randomized model than for the strict 
cut-in-two
version, for which $\tilde p(z)$ is known to decay exponentially for large $z$
\cite{Godreche}. Note however that the density of small clusters strictly 
vanishes
below a certain value $z_c$ in the cut-in-two model, whereas it is finite in 
the
present dynamics. This is due to the fact that although small clusters are 
typically
the first ones to disappear, fluctuations can persist and keep some small 
clusters
alive, whereas by definition they systematically disappear with a cut-in-two 
rule.
The excess
density of small clusters has the interesting consequence, discussed below, to
give rise to a sub-aging behaviour.

\begin{figure}[htbp]
\begin{center}
\centerline{\psfig{figure=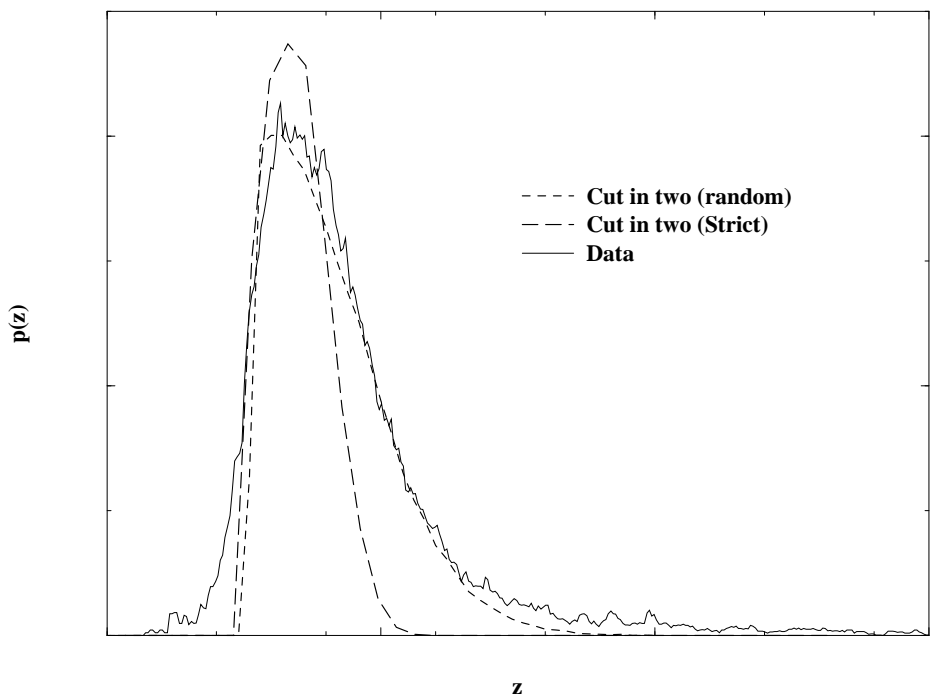,width=8cm,angle=270}}
\caption{Comparison of the distribution of the cluster sizes in our model
with the ones corresponding to a strict cut-in-two model
and a randomized cut-in-two model. Note that the density of small clusters
strictly vanishes in the latter two models.}
\end{center}
\end{figure}

\subsection{Aging at long times}

Another manifestation of the long lived metastable states is the aging
phenomenon, which we have investigated numerically by computing the
following age dependent {\it structure factor} \cite{MB}:
\be
C_{q}(t_w,t+t_w)=
\frac{1}{N} \sum_{i=1}^N \langle e^{iq(x_i(t+t_w)-x_i(t_w))} \rangle
\ee
This quantity is expected to depend both on $t_w$ and $t$ (violation of
time translation invariance) in the out-of-equilibrium regime where 
$T \log t_w \ll N$. As can be seen from Figure 127, obtained for $T=5$, 
$N=10000$ and
averaged over $10$ initial configurations, the relaxation is slower 
and slower as $t_w$ increases. As often observed experimentally
\cite{Sitges,BCKM,StA}, we  have
found that the data can be rescaled in the variable $s=t/t_w^\mu$, with $\mu 
<1$ (subaging). Within the time regime probed by the {\sc mc},
$\mu \sim 0.75$ for $q=5$ and $q=10$. This means that 
the characteristic relaxation time grows more
slowly that the waiting time itself. In the present situation, this can
be interpreted following the above remark about the small $z$ behaviour
of $\tilde p(z)$: the presence of clusters much smaller than the average size
$\overline n$ means that some clusters will evolve on a time scale much shorter
than $t_w=\exp(\overline n/T)$. More precisely, one can approximate the
short time behaviour of $C_q(t+t_w,t_w)$ by:
\be
1 - C_q(t+t_w,t_w) \simeq t \int dz \ z \tilde p(z) \exp(-z \overline n/T).
\ee
The above equation is obtained by assuming that each cluster of size
$n$ contributes to the correlation function as $n\exp(-t/\tau(n))$
where $\tau(n)=\exp(n/T)$.
If $\ln \tilde p(z)$ behaves as $(z_0/z)^{c}$ for small $z$, a saddle-point 
approximation of the above integral leads to:
\be
1 - C_q(t+t_w,t_w) \propto \frac{t}{t_w^\mu} \qquad \mu=(1+\frac{1}{c})
\left(\frac{c z_0^c}{\ln t_w}\right)^{\frac{1}{1+c}}.\label{muscaling}
\ee
This scenario therefore leads to a 
subaging
behaviour for large $t_w$, albeit with a (slowly) time dependent exponent 
$\mu$. The
absence of
`small' clusters would correspond to the limit $c \to \infty$, 
such that $\tilde p(z < 1)=0$. In this case, one finds $\mu=1$ as expected. 
The rescaled function $C_q(s)$ can
be fitted by $C/(s_0+s)^w$, with $w \simeq 0.18$ for $q=5$ and $w \simeq 0.10$ 
for $q=10$.

The above mechanism for subaging might be much more general, and hold for 
other slowly
coarsening systems, such as the Random Field Ising model. Suppose that the 
distribution
of domain sizes scales with an age dependent average length $\overline 
R(t_w)$, such that
$t_w \sim \exp(\overline R^\psi/T)$ \cite{RFIM}. Then the presence of domains smaller than 
$\overline R(t_w)$ will correspond to relaxation times much shorter than 
$t_w$, and
therefore to the possibility of subaging effects.

\begin{figure}[htbp]
\begin{center}
\centerline{\psfig{figure=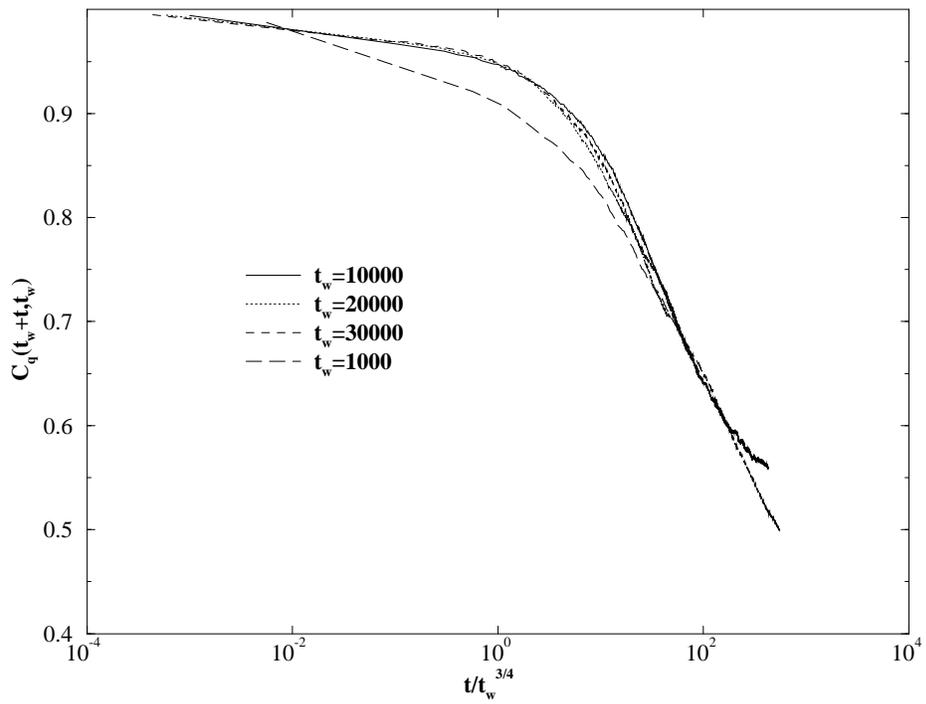,width=8cm,angle=270}}
\caption{Evidence for aging in the structure factor,
for $N=10000$, $q=5$, $T=5$, averaged over 10 initial conditions.
The curves for $t_w=10^4, 2\ 10^4$ and $3\ 10^4$ superimpose when plotted
as a function of $t/t_w^{3/4}$. The curve with $t_w=1000$, however, has a
different shape. The mechanism proposed in the text suggests that the scaling
in $t/t_w^\mu$ should indeed only be approximate.}
\end{center}
\end{figure}

\section{Summary -- Conclusion}

We have investigated a simple one dimensional model for the swelling or
the collapse of a closed polymer chain of size $N$, representing the dynamical 
evolution
of a polymer in a $\Theta$-solvent that is rapidly changed into a good 
solvent (swelling) or a bad solvent (collapse).
\begin{itemize}
\item
In the case of swelling, 
the initial phase is to a large degree independent of the chain constraint.
The density profile is parabolic and expands in space as $t^{1/3}$. This regime
is well described by a Flory-like continuum equation for the monomer density.
The dynamics slows down after a time $\propto N^2$ when the chain becomes 
stretched,
and the polymer gets stuck in metastable `zig-zag' configurations 
(that we have enumerated), from which it escapes through thermal activation. 
The final stage of the stretching is therefore found to be logarithmic in 
time:
the size of the polymer grows as $\sqrt{\ln t}$.
\item In the case of collapse, the chain very quickly (after a time of order 
unity)
breaks up into small `pearls' or clusters of monomers. The evolution of the 
chain
towards the
fully collapsed states proceeds through a slow growth of the size of these
metastable clusters, again evolving as the logarithm of time. We have 
enumerated
the total number of metastable states as a function of the extension of the 
chain,
and deduced from this computation that the radius of the chain should 
decrease as a double logarithm of time, i.e. $1/\ln(\ln t)$. We have also 
found
the total number of metastable states with a given value of the energy. We 
also
obtained
the distribution of cluster sizes, that we compared to simple one dimensional 
coalescence models. Finally, the aging properties of the dynamical structure 
factor was investigated numerically. The subaging behaviour that we found was
attributed to the tail of the distribution at small cluster sizes, 
corresponding
to anomalously fast clusters. We have argued that this mechanism for subaging 
might
hold in other slowly coarsening systems.
\end{itemize}

The system studied here falls in the category of non disordered, finite
dimensional models exhibiting glassy dynamics. The list of such models keeps
increasing with time \cite{Othermodels}, although the present model is quite 
realistic (no ad-hoc dynamical rules or long-ranged interaction). 
Actually the chain constraint does not play a very important role for the
dynamics of collapse which is, as mentioned above, quite close to the
one-dimensional version of the Backgammon model. 

In real polymeric systems, the presence of a chain would change quantitative 
features
such as diffusion coefficients. Polymer melts have been shown 
to behave very similarly to structural glasses such as binary mixtures of 
Lennard-Jones
particles \cite{Baschnagel}. These polymeric systems are out of equilibrium at 
long times
and attempts have been made to describe them within 
the framework of the Mode Coupling Theory \cite{Baschnagel,Shakh}. 

A limitation of the present model in the collapsing regime is the unrealistic
nature of the interaction potential: there is no hard-core constraint, and
the range of the attractive part is finite. For realistic potentials, the
total energy of a cluster containing $n$ monomers will eventually grow like 
$n$
for large $n$ and not like $n^2$. Correspondingly, the energy barrier to remove
one monomer from a large cluster is finite, instead of being of order $n$. 
Furthermore, the long-ranged
part of the attractive potential can speed up the dynamics by making the large
clusters move towards each other, an effect absent in the present model. 
Nevertheless, the model used here could be justified for colloidal particles,
of the type studied in \cite{Colloids}. In this system, clusters of particles
form and slowly coalesce, this leads to a very clear aging of the
dynamical structure factor. Hints about the presence of `micro-earthquakes',
possibly due to the disappearance of large clusters, have been reported. A 
closer comparison between our model and such experiments would be very 
fruitful,
but might require the inclusion of elastic deformations, as suggested in 
\cite{Colloids}.

\section*{Acknowledgments}
We would like  to thank I. Campbell, C. Godr\`eche, F. Ritort,
D. Sherrington, P.
Sollich and J. Vannimenus for helpful discussions. E. I. Shakhnovich must be
acknowledged for his hospitality while part of this work was done.
 EP was
supported  by NIH grant GM 52126 
and by Centre National de la Recherche Scientifique (France).

\section*{Appendix A: A variational method for the swelling of a polymer chain}

In \cite{Orland} a polymeric chain
with given two-body interactions is approximated by a `variational' gaussian 
chain
with a time-dependent Kuhn length.
In other words one starts from the original chain with hamiltonian $H$ and
its associated Langevin equation:
\begin{eqnarray}
    && \frac{\partial{r}}{\partial{t}} = - \Gamma_0
        \frac{\partial{H}}{\partial{r}} + \eta(s,t) \label{1}\\
    && H=\frac{k_B T}{2 a_{0}^{2}} \int_{0}^{N}
        \left(\frac{\partial{r}}{\partial{s}}\right)^{2}ds 
              + k_B T V(r(s,t)) \label{2}
\end{eqnarray}
and then one tries to replace it with another system characterized by $H_v$:
\begin{eqnarray}
    && \frac{\partial{r^{(v)}}}{\partial{t}} = -\Gamma_0
        \frac{\partial{H_{v}}}{\partial{r^{(v)}}} + \eta(s,t) \label{3}\\
    && H_{v}=\frac{ k_B T}{2 a^{2}(t)} \int_{0}^{N}
       \left (\frac{\partial{r^{(v)}}}{\partial{s}}\right)^{2}ds.  \label{4}
\end{eqnarray}
$\eta(s,t)$ is a gaussian noise with
\be
\langle \eta(s,t)\eta(s,t') \rangle=2 D \delta(t-t')
\ee
and
$\Gamma_0={D \over k_B T }$ where $D$ is the diffusion coefficient.

In order to find the effective Kuhn length that would represent best
the initial system,
the radius of gyration is chosen to coincide in both models,
to first order in  $ \chi(s,t)=r(s,t)-r^{(v)}(s,t)$ and
$W= H - H_v$.

This leads to an implicit equation for $a(t)$:

\be
   \int_{0}^{N}<r^{(v)}(s,t)\chi(s,t)>=0 \label{8}
\ee

We have used periodic boundary conditions, introducing
$\omega_{n}=\frac{2\pi n}{N}$.

The whole calculation is discussed in detail in \cite{Orland}, and we only 
reproduce
the final equations which appear in the form $I+II=0$, with,
in the case of two-body interactions:
\be
I=-\sum_{n\neq 0} \omega_n^2
\left[\frac{a_0^2}{N\omega_n^2} e^{-2D\omega_n^2 f(t)}
(\frac{t}{a_0^2} -f(t))
+\frac{2D}{N} \int_0^t d\tau
(\frac{1}{a_0^2}-\frac{1}{a^2(\tau)}) 
\int_0^{\tau} d\lambda e^{-2D\omega_n^2 (f(t)-f(\lambda))} \right]
\ee
and
\be
II=2vN \sum_{n\geq 1} e^{-2D\omega_n^2 f(t)}
\int_0^1 dU \int_0^t d\tau 
\frac{1-\cos 2\pi n U}{\alpha(\tau, U)^{1+\frac{d}{2}}}
\left[\frac{a_0^2}{N\omega_n^2}
+ \frac{2D}{N} \int_0^{\tau} d\lambda e^{2D\omega_n^2 f(\lambda)}
\right].
\ee
We have set:
\be
\alpha(\tau, U)=4 \sum_{p\geq 1} (1-\cos 2p\pi U) 
\left[\frac{a_0^2}{N\omega_p^2}e^{-2D\omega_p^2 f(\tau)}
+\frac{2D}{N} \int_0^{\tau} d\lambda
e^{-2D\omega_n^2 (f(\tau)-f(\lambda))}
\right],
\ee
and $f(t)=\int_0^t \frac{d\tau}{a^2(\tau)}$.

We are interested here in an intermediate time regime, and choose to keep the 
time
dependence in $a(t)$ such that
$R(t)=a(t)N^{\frac{1}{2}}\ll N^{\frac{3}{d+2}}$
and look for a power-law scaling:
$a(t)=N^{\mu} t^{\alpha}$. Then the reduced variable
$u=\frac{t}{N^{\gamma}}$
with $\gamma=\frac{1}{\alpha} [\frac{4-d}{2(d+2)}-\mu]$
is the relevant parameter in this time regime; we are interested in the 
situation
where one can take the limit $N\rightarrow\infty$, keeping $u$ finite and much 
smaller
than unity.

A succession of approximations gives the following behaviours for the 
quantities $I$ and $II$:
\be
I \simeq u^{2 -\frac{3x}{2}} N^{2\gamma +3\mu-\frac{3}{2} x \gamma}
\qquad II \simeq u^{1-(1-\frac{x}{2})(1+\frac{d}{2})}
N^{2+\gamma -\rho (1+\frac{d}{2})} \qquad \alpha(\tau, U) \simeq N^{\rho} 
v^{1-\frac{x}{2}},
\ee
where the parameters used above are defined as:
\be
x=1-2\alpha,\qquad \rho=\gamma+\mu -\frac{1}{2} x \gamma,\qquad 
v=\frac{\tau}{N^{\gamma}}.
\ee
Finally this allows to find the result announced in the main text
for the case of repulsive interactions:
\be
\alpha=\frac{1}{2} \frac{2-d}{6-d} \qquad (d < 2),
\ee
so that, in $d=1$, $R(t) \simeq t^{\frac{1}{14}}  N^{\frac{11}{14}}$.

\section*{Appendix B: Calculation of the number of metastable states in the 1D 
Kawasaki
Ising model}

Cornell, Kaski and Stinchcombe 
\cite{Kawasaki} have studied the dynamics of the one dimensional Ising model, 
using the Kawasaki prescription: the total magnetization is fixed and the only 
moves allowed are exchanges of two antiparallel nearest-neighbour spins. It is 
best suited to model binary alloys. At low temperature, the model has been 
shown
to exhibit a glassy-type behaviour.
At finite temperature, the system evolves in time by condensation of domains 
of
parallel spins (if $J$ is the exchange coupling constant, there is an energy 
gain
of $4J$ for a spin to join a domain by switching sites with its nearest 
neighbour);
conversely, the evaporation of a spin from a domain costs $4J$. The 
diffusion of a spin performing a random walk in a domain consisting of spins 
of
opposite sign costs no energy.

At zero temperature, all evaporation processes are frozen and only 
condensation
and diffusion of spins can occur.
However, in order for two domains of same sign to coalesce,
the evaporation of the boundary spins is necessary;
as a consequence, at $T=0$, many states are frozen.

The frozen states are in fact very easy to characterize:
they consist of alternating positive and negative domains, each of them 
containing at
least two spins.
Let $M$ be the total number of positive domains;
for $i=1,\dots, M$, $n_i^+$ is the number of $+$ spins in the positive domain 
$i$.
Between two of these domains there exists a domain of negative magnetization.
We label also these domains by $i=1,\dots, M$, and $n_i^-$ is the number of
$-$ spins in the negative domain $i$.

Since $N$ is the total number of spins and $Nb$ the total magnetization (the
letter $m$ being used in the text for the chain constraint),
we have the two conservation relations:
\be
N=\sum_{i=1}^M (n_i^+ +n_i^-); \qquad Nb=\sum_{i=1}^M (n_i^+ -n_i^-).
\ee
The number of frozen states for $M$ fixed and fixed magnetization 
is given by the following expression:
\be
N_F(M,b)=(\sum_{n_1^+=2}^N \dots \sum_{n_M^+=2}^N )
(\sum_{n_1^-=2}^N \dots \sum_{n_M^-=2}^N)
\delta\left(\sum_{i=1}^M (n_i^+ +n_i^-)-N\right)
\delta\left(\sum_{i=1}^M (n_i^+ -n_i^-)-Nb\right) \label{kawasaki}.
\ee
The calculation can be handled easily by exponentiating the 
$\delta$-functions,
and taking the $N\rightarrow \infty$ limit.

\begin{eqnarray*}
N_F(M,b)&=&
\int d\lambda e^{-i\lambda N}
\int d\mu e^{-i \mu Nb}
\sum_{n_1^+=2}^N \dots \sum_{n_M^+=2}^N 
e^{(i\lambda +i\mu)\sum_{i=1}^M n_i^+}
\sum_{n_1^-=2}^N \dots \sum_{n_M^-=2}^N
e^{(i\lambda -i\mu)\sum_{i=1}^M n_i^-}\\
&=&
\int d\lambda e^{-i\lambda N}
\int d\mu e^{-i \mu Nb}
\left(\frac{e^{2i(\lambda+\mu)}}{1-e^{i(\lambda+\mu)}}\right)^M
\left(\frac{e^{2i(\lambda-\mu)}}{1-e^{i(\lambda-\mu)}}\right)^M
\end{eqnarray*}
Then a saddle point approximation on $z_1=i\lambda$ and $z_2=i\mu$
leads to  a simple expression for the entropy of frozen states at fixed
magnetization per spin $m$ and fixed fraction of domains 
$\delta=\frac{2M}{N}$.

\begin{eqnarray*}
\frac{1}{N}\ln N_F(M,b)&=&S(b,\delta)=\frac{1}{2} (-1-b+2\delta)
\ln(1+b-2\delta) +\frac{1}{2} (1+b-\delta) \ln(1+b-\delta)\\
&+&\frac{1}{2} (-1+b+2\delta) \ln(1-b-2\delta)
+\frac{1}{2} (1-b-\delta) \ln(1-b-\delta)
-\delta \ln \delta.
\end{eqnarray*}

We then determine the optimum $\delta^*$
for $\delta$ at fixed magnetization $b$, in order to find the total entropy
for a given $b$. This leads to an implicit equation:

\be
\frac{(1+b-2\delta^*)^2(1-b-2\delta^*)^2}{\delta^{*2}
(1+b-\delta^*)(1-b-\delta^*)}=1
\ee

\subsection*{Case $b=0$}

The complexity of the Kawasaki model is then easy to compute,

\be
{\cal S}^{K}(b=0)=(-1+2\delta) \ln(1-2\delta)
+(1-\delta) \ln(1-\delta)
-\delta \ln \delta,
\ee

The sum over $\delta$ can be performed
using  a saddle-point approximation, leading to a total number of frozen
states:
\be
N_F ^{K}(b=0)\sim \frac{C^N}{\sqrt{N}},
\ee
where the complexity per spin is $C\simeq 1.618$. This is slightly less than the
number of closed random walks for which $C=2$: each metastable
configuration in the Kawasaki problem can be seen as a random
walk that goes at least twice in the same direction before changing
direction.

The case of zero magnetization is of special interest 
because it allows to give an estimate of the complexity of the polymeric chain
in its stretching regime.
As was pointed out in section $3.3$,
as the chain swells,
it is trapped  in metastable
configurations, consisting of fully stretched segments of the chain going
alternatively to the left or to the right. It is easy to draw a correspondence
between a segment going to the right (respectively, to the left)
and a domain of plus (respectively, minus) spins. The constraint that the 
chain is closed
requires that
the number of steps to the right is equal to the number of steps to the left,
which is equivalent, in the spin language, to $b=0$.

In order to count the {\it total} number of metastable states for the chain,
one should actually require that each stretched segment must contain at least
one monomer, so that the sums in (\ref{kawasaki}) should start at
$n_i^{\pm}=1$. However, there must be an energetic bias in order for each end 
of
segment to be stuck at its position, namely that the number of monomers on its
`inward' neighbouring site along the chain should be more important. As
explained in  the main text, we assume that
this is true with probability $1/2$, and this adds a multiplicative factor
$\frac{1}{2^M}$ in formula (\ref{kawasaki}). Finally the complexity of the
chain is:
\be 
{\cal S}=-\delta \ln 2 +(\delta-1) \ln(1-\delta)  -\delta
\ln \delta,
\ee

This relation is illustrated on Fig. 4
as a function of $R$, given that $R\sim M^{-\frac{1}{2}}$.
The sum over $\delta$ can be performed
using  a saddle-point approximation, leading to a total number of frozen
states $N_F \sim {C^N}/{\sqrt{N}},$ where $C=3/2$.

To summarize, the calculation reveals an interesting analogy:
the number of states of a chain containing at least $k$ monomers on
each stretched segment is the same as the number of the frozen states in the
1D Kawasaki spin model at zero magnetization and zero temperature
(with all domains
containing at least $k$ spins), and is in turn equal to the number of
one-dimensional closed random
walks biased in such a way that one has to walk at least $k$ times in the same
direction before changing direction.

In order to know the number of {\it metastable} states of the chain with at
least $k$ monomers on each segment, one must add an energetic bias introduced
via the additional factor $\frac{1}{2^M}$ in equation (\ref{kawasaki}).

To complete the analogy let us formulate two remarks.

First, the number of closed random walks in one dimension
($2^N/\sqrt{N}$) is
recovered when one  uses the above formula (\ref{kawasaki}) with sums starting
at $n_i^{\pm}=1$.

Second, one can also vary the degree of bias $k$ of the random walk
by starting the sums in (\ref{kawasaki}) at $n_i^{\pm}=k$. Physically this 
corresponds
to metastable states over a time scale $t$ such that $\ln t =\beta v k$. If in 
addition,
we incorporate 
the energetic bias factor $\frac{1}{2^M}$, this allows to follow
with  time
the number of metastable states at finite temperature as the system coarsens,
and relate it to the radius of gyration of the chain at long times.

We obtain the following entropy of metastable states, as a function of the
parameter $u=T \ln t$:

\be
{\cal S}(u)=-\delta \ln 2 +(-1+u \delta) \ln(1-u \delta)
+(1-(u-1)\delta) \ln(1-(u-1)\delta)
-\delta \ln \delta,
\ee

This entropy is maximum at a certain value of $\delta$, which, to leading
order in $u$ (for $u$ large), is given by $\delta^* \simeq
\frac{1}{u}$. We expect that the number of zig-zags in the polymer will adjust 
at this
value, corresponding to one of its most probable configurations.
As was explained in
the main text, $R$ and $\delta^*$ are then related  by
$\delta \sim N/R^2$, so that $R \sim \sqrt{N \ln t}$.

Finally,
let us mention that the whole calculation
is in spirit (and also in the
analytical methods involved) very similar to
the calculation of the number of metastable states of a collapsing chain,
as a function of the degree of freezing $k^*=T\ln t$; section $4$ of this
article is devoted to such a description.

\section*{Appendix C: Number of  collapsed metastable states as a function of
the  energy}

We start by the expression of the 
number of $k^*$-stable configurations with $M$ clusters
and $N$ monomers, at fixed energy E, given by:
\be
{\cal N}_{\rho,k^*} (E,M|N) = \sum_{n_1=k^*}^{N}
\dots \sum_{n_M=k^*}^{N} \delta\left(\sum_{i=1}^{M} (n_i-2)+2L -N\right)
\delta\left(\sum_{i=1}^{M} -\frac{1}{2} n_i(n_i-1)-(L-M)-E\right).
\ee

In the following, we set $\epsilon=\frac{E}{N}<0$ and limit ourselves to the
case $m=1$.

Let us first note that for a given $k^*$, the maximum energy per particle
is equal to $\epsilon_{\max}=-\frac{1}{2} (k^*+1)+\frac{k^*}{\rho}$. It
is reached for a configuration consisting of $M$
clusters of $k^*$ monomers each, each of energy $-k^* (k^*-1)/2$
 and the remaining $L-M$ sites are occupied by $2$ monomers, each of these
sites contributing with an energy equal to $-1$. The conservation relation
$N=Mk^* +2(L-M)$ leads to 
$\delta_{max}=\frac{M_{max}}{N}=(1-2/\rho)/(k^*-2)$, and the energy is
calculated accordingly, leading to
$\epsilon_{\max}=-\frac{1}{2} (k^*+1)+\frac{k^*}{\rho}$.
In the following, we will only need to
consider values of $\epsilon$ in the interval $]-\infty;
\epsilon_{\max}]$.

From the expression of ${\cal N}_{\rho,k^*} (E,M|N)$ above,
one may use the integral representation of the $\delta$-functions to find that:
\be
{\cal N}_{\rho,k^*} (E,M|N) = \int dX \int dY e^{-N F(X,Y)}.
\ee
with 
\be
F(X,Y)= X(1-\frac{2}{\rho})+Y(\epsilon+\frac{1}{\rho}) - \delta \ln
\Sigma(X,Y), \ee
and
\be
\Sigma(X,Y)= \sum_{n=k^*-2}^{\infty} e^{Xn -\frac{1}{2}Yn(n+3)}.\ee

Our procedure is to find an estimate of this quantity via a saddle-point 
approximation,
so that the entropy of metastable states at fixed $\delta$ is given by:
\be
S_{\rho}(\delta, \epsilon)=-F(X^*,Y^*)
-\delta \ln\left(\frac{\delta\rho(1-\delta \rho)}{(1-2\delta \rho)^2}\right)
+\frac{1}{\rho} \ln\left(\frac{1-\delta \rho}{1-2\delta \rho}\right)
\ee
where the last step is to maximize this expression with respect to $\delta$
so that the final result is $S_{\rho}(\epsilon)= S_{\rho}(\delta^*, \epsilon)$.

\subsection*{Limit of low energies $\epsilon \rightarrow -\infty$}

In this part, it is more convenient to perform a change in variables
$Z=X-3Y/2$ and $Z'=-{Y}/{2}$. Moreover, numerical investigation
suggests that for $|\epsilon|$ large, $Z^*$ should be small and $Z'^*$ of the 
form
$Z'^*\simeq -a Z^{*2}, a>0$; so we set in the following
$Z'\simeq -a Z^2, a>0$, so $Z$ and $a$ are the new variables:
\be
F(Z,a)=Z(1-\frac{2}{\rho})+aZ^2(3+2\epsilon-\frac{4}{\rho}) -\delta \ln
\Sigma(Z,a), \ee
and:
\be
\Sigma(Z,a)= \sum_{n=k^*-2}^{\infty} e^{-a Z^2 n^2 +Zn}
\ee

We then approximate the discrete sum $\Sigma(Z,a)$ by an integral, which is 
valid
if we find {\it a posteriori} that $Z$ is small:
\be
\Sigma(Z,a) \simeq -\frac{1}{Z} G(a), \ \
  G(a)= \int_{0}^{\infty} du e^{-au^2-u}
\ee
Up to this order in ${1}/{Z}$ the results are independent of $k^*$.
The combined saddle-point equations
${\partial F}/{\partial Z}=0$ and
${\partial F}/{\partial a}=0$ lead to the simple relation:
\be
Z^*(1-\frac{2}{\rho})=-\delta -2a^* \delta \frac{G'(a^*)}{G(a^*)}.
\ee
Continuing the algebra with the assumption that $a^* \neq 0$
does lead to inconsistent values for $\delta^*$ ($\delta^*>\frac{1}{2}$).
We conclude that the saddle-point solution, if there is one, is stuck
at $a^*=0$. The rest of the calculation is then straightforward:
we obtain $Z^*=-\delta/(1-\frac{2}{\rho})$, and since $G(0)=1$,
\be
S_{\rho}(\delta, \epsilon)=\delta[1+\ln(1-\frac{2}{\rho})] -\delta \ln \delta
-\delta \ln\left(\frac{\delta\rho(1-\delta \rho)}{(1-2\delta \rho)^2}\right)
+\frac{1}{\rho} \ln\left(\frac{1-\delta \rho}{1-2\delta \rho}\right)
\ee
Finally, in order to determine $\delta^*$, we compute:
\be
\frac{\partial S_{\rho}(\delta, \epsilon)}{\partial \delta}=
\ln(1-\frac{2}{\rho})-\ln \left[\frac{\delta^2 \rho (1-\delta \rho)}{(1-2\delta
\rho)^2}\right] \ee
In the limit $\epsilon \rightarrow -\infty$, the interval allowed for 
$\delta$ is $[0; \delta_{\max}= -(1-\frac{2}{\rho})^2/2\epsilon]$. In this
interval,  $S_{\rho}(\delta, \epsilon)$ is an increasing function of $\delta$,
and therefore  reaches its maximum at the upper bound. Therefore:
$\delta^*=\delta_{\max}$.  Note that this implies that
$Z^*={(1-\frac{2}{\rho})}/{2\epsilon}$ is indeed small as was assumed  in this
approach. We get the final result for the tail of the entropy of metastable 
states : \be
S_{\rho}(\epsilon) \simeq_{\epsilon \to -\infty}
-\frac{\ln(-\epsilon)}{\epsilon} (1-\frac{2}{\rho})^2
+\frac{1}{\epsilon}(1-\frac{2}{\rho})^2 [-1+\frac{1}{2}
\ln(\frac{\rho}{4}(1-\frac{2}{\rho})^{3/2})] + o(\frac{1}{\epsilon^2})
\ee

\subsection*{Limit of high energies $\epsilon \rightarrow \epsilon_{\max}$.}

We start again from the expression for the free energy in terms of the 
variables
$X$ and $Y$. We assume that $Y^*$ to be large, and a good 
approximation is
to keep only the first two terms in the sum $\Sigma(X,Y)$,
corresponding to $n=k^*-2$ and $n=k^*-1$. Then,
\be
F(X,Y)= X (1-\frac{2}{\rho}-\delta(k^*-2)) + Y (\epsilon+
\frac{1}{\rho}+\frac{1}{2} \delta (k^*-2)(k^*+1))   - \delta \ln (1+e^{X-k^*
Y}). \ee
The saddle-point equations ${\partial F}/{\partial X}=0$
and ${\partial F}/{\partial Y}=0$ lead to a simple system for
$x=e^X$ and $y=e^Y$:
\be
\frac{x}{y^{k^*}}=\frac{1-\frac{2}{\rho}-\delta(k^*-2)}{\delta
(k^*-1)-1+\frac{2}{\rho}};\ \
\frac{x}{y^{k^*}}=-\frac{\epsilon+\frac{1}{\rho}+\frac{1}{2}\delta(k^*-2)(k^*+1
)}
{\epsilon+\frac{1}{\rho}+\frac{1}{2}\delta(k^*-1)(k^*+2)}. \ee
From this system we obtain a relation between $X$ and $Y$:
\be
X=k^* Y + \ln \left( \frac{1-\frac{2}{\rho}-\delta(k^*-2)}{\delta
(k^*-1)-1-\frac{2}{\rho}} \right), \ee
as well as a consistency relation between $\delta$ and $\epsilon$:
\be
\delta^*=2\frac{\epsilon+\frac{1}{\rho}+k^*(1-\frac{2}{\rho})}{(k^*-1)(k^*-2)}
\ee
Let us note that at the upper boundary of the interval studied, i.e. for 
$\epsilon=\epsilon_{max}=- \frac{1}{2}(k^*+1)+\frac{k^*}{\rho}$,
this consistency relation gives $\delta=\delta_{max}=(1-2/\rho)/(k^*-2)$,
which is indeed the exact result.
Then the expression for entropy can be computed in a simple way:
\bea\nonumber
S_{\rho}(\epsilon)&=&
-(1-\frac{2}{\rho}-\delta^*(k^*-2)) 
\ln \left( \frac{1-\frac{2}{\rho}-\delta^*(k^*-2)}
{\delta^*(k^*-1) -1+\frac{2}{\rho}} \right) + \delta^* \ln
\left(\frac{\delta^*}{\delta^*(k^*-1) -1+\frac{2}{\rho}} \right)\\
 &-&\delta^* \ln\left(\frac{\delta^*\rho(1-\delta^*
\rho)}{(1-2\delta^* \rho)^2}\right) +\frac{1}{\rho} \ln\left(\frac{1-\delta^*
\rho}{1-2\delta^* \rho}\right) \eea
This gives an estimate for the entropy of metastable states near 
$\epsilon_{max}(k^*)$
for a given $k^*$. Note that the expression found above is actually defined 
only in the
interval $[\epsilon_{max}(k^*+1)=- \frac{1}{2}(k^*+2)+\frac{k^*+1}{\rho};
\epsilon_{max}(k^*)=-\frac{1}{2}(k^*+1)+\frac{k^*}{\rho}]$, beyond which
the present approximation breaks down. 

To summarize, for each $k^*$, or equivalently for each $u=T \ln t$,
we found the behaviour of the entropy of metastable states both for very low 
energies
$(\epsilon \rightarrow -\infty)$, where it is independent of $k^*$, and in the 
vicinity of
the maximum energy $\epsilon_{max}(k^*)$. Thus the full curve for
$S_{\rho}(\epsilon, k^*)$ is an extrapolation between these two asymptotic 
regimes,
as shown in Fig. 10.

\bibliographystyle{unsrt}

\end{document}